\documentclass[fleqn,usenatbib]{mnras}

\usepackage{newtxtext,newtxmath}

\usepackage[T1]{fontenc}
\usepackage{ae,aecompl}

\usepackage{graphicx}	
\usepackage{amsmath}	
\usepackage{amssymb}	
\usepackage{multirow}
\usepackage{booktabs}
\usepackage{gensymb}
\usepackage{siunitx}
\usepackage{multirow}
\usepackage{array}
\usepackage{multicol}
\usepackage[mathlines]{lineno}

\usepackage{etoolbox} 

\newcommand*\linenomathpatch[1]{%
  \expandafter\pretocmd\csname #1\endcsname {\linenomath}{}{}%
  \expandafter\pretocmd\csname #1*\endcsname{\linenomath}{}{}%
  \expandafter\apptocmd\csname end#1\endcsname {\endlinenomath}{}{}%
  \expandafter\apptocmd\csname end#1*\endcsname{\endlinenomath}{}{}%
}
\newcommand*\linenomathpatchAMS[1]{%
  \expandafter\pretocmd\csname #1\endcsname {\linenomathAMS}{}{}%
  \expandafter\pretocmd\csname #1*\endcsname{\linenomathAMS}{}{}%
  \expandafter\apptocmd\csname end#1\endcsname {\endlinenomath}{}{}%
  \expandafter\apptocmd\csname end#1*\endcsname{\endlinenomath}{}{}%
}

\title[Constraining Models of the PWN in SNR G0.9+0.1 with CTA]{Constraining Models of the Pulsar Wind Nebula in SNR G0.9+0.1 via Simulation of its Detection Properties using the Cherenkov Telescope Array}
\author[Fiori, Zampieri, Burtovoi, et al.]{
M. Fiori,$^{1,2}$ \thanks{E-mail: michele.fiori86@gmail.com}
L. Zampieri,$^{2}$
A. Burtovoi,$^{2,3}$
P. Caraveo,$^{4}$
and L. Tibaldo$^{5}$
\\
$^{1}$Department of Physics and Astronomy, University of Padova, Via F. Marzolo 8, I-35131, Padova, Italy\\
$^{2}$INAF-Osservatorio Astronomico di Padova, Vicolo dell'Osservatorio 5, I-35122, Padova, Italy\\
$^{3}$Centre of Studies and Activities for Space (CISAS) 'G. Colombo', University of Padova, Via Venezia 15, I-35131 Padova, Italy\\
$^{4}$INAF-IASF Milano, Via A. Corti 12, I-20133 Milano, Italy\\
$^5$IRAP, Universit\'e de Toulouse, CNRS, CNES, UPS, 9 avenue Colonel Roche, 31028 Toulouse, Cedex 4, France}

\date{Accepted 29/09/2020}

\pubyear{2020}

\begin{document}

\label{firstpage}

\pagerange{\pageref{firstpage}--\pageref{lastpage}}

\maketitle

\begin{abstract}
SNR G0.9+0.1 is a well known source in the direction of the Galactic Center composed by a Supernova Remnant (SNR) and a Pulsar Wind Nebula (PWN) in the core. We investigate the potential of the future Cherenkov Telescope Array (CTA), simulating observations of SNR G0.9+0.1. We studied the spatial and spectral properties of this source and estimated the systematic errors of these measurements. The source will be resolved if the VHE emission region is bigger than $\sim0.65'$. It will also be possible to distinguish between different spectral models and calculate the cut-off energy. The systematic errors are dominated by the IRF instrumental uncertainties, especially at low energies. We computed the evolution of a young PWN inside a SNR using a one-zone time-dependent leptonic model. We applied the model to the simulated CTA data and found that it will be possible to accurately measure the cut-off energy of the $\gamma$-ray spectrum. Fitting of the multiwavelength spectrum will allow us to constrain also the magnetization of the PWN. Conversely, a pure power law spectrum would rule out this model. Finally, we checked the impact of the spectral shape and the energy density of the Inter-Stellar Radiation Fields (ISRFs) on the estimate of the parameters of the PWN, finding that they are not significantly affected.
\end{abstract}

\begin{keywords}
cosmic rays - gamma-rays - pulsars - supernova remnants 
\end{keywords}

\section{Introduction}

Pulsar Wind Nebulae (PWNe) represent the most numerous class of identified Galactic Very High Energy (VHE)  $\gamma$-ray sources \citep{deOna2013pwne}. These objects are highly magnetized nebulae powered by young and energetic pulsars.
Inside these nebulae non-thermal radiation up to $\sim100$ TeV is produced \citep{2013FrPhy...8..714R}.

In young PWNe the outer radius of the nebula has not yet started to interact with the reverse shock of the SNR. Therefore, they are particularly interesting objects because the uncertainties related to the interaction are not present and their evolution can be fairly well reproduced by physical models. These models can thus be tested against observations, providing important information on the physical processes at work in these sources \citep[e.g.][]{gelfand2009, martin2012, zhu2015}.

The Cherenkov Telescope array \citep[CTA,][]{cta2011} will be capable to study the $\gamma$-ray emission of PWNe in great detail.
With CTA it will be possible to observe PWNe from few GeV up to hundreds of TeV, accurately sampling most of the Inverse-Compton (IC) peak as well as obtaining a measurement of the spectral cut-off energies where present. In addition, the unprecedented angular resolution will allow us to determine more precisely the $\gamma$-ray emission regions and to investigate the existence of any potential energy-dependent morphology. Thanks to this it will be possible to test various $\gamma$-ray emission models of PWNe and to better understand their magneto-hydrodynamic structure and evolution.

The purpose of this work is testing the capabilities of CTA in connection with a specific source (SNR G0.9+0.1) while, at the same time, assessing the impact of CTA observations on our understanding of the physical processes occurring in PWNe.
The source selected is SNR G0.9+0.1 (at TeV energies the source is also referred as HESS J1747-281; \citealt{HessGPS2018}), a well known composite Supernova Remnant \citep[SNR,][]{1987ApJ...314..203H}. 
The bright central core has been unambiguously identified as a PWN through X-ray observations \citep{2001ApJ...556L.107G}.
SNR G0.9+0.1 is composed by a PWN in the core (with a diameter of $\sim2'$) surrounded by a SNR \citep[with a diameter of $\sim8'$, ][]{2008A&A...487.1033D}. This source has been detected at VHE by HESS \citep{2005A&A...432L..25A}, VERITAS \citep{2015arXiv150806311S} and MAGIC \citep{2017A&A...601A..33A} only up to $\sim20$ TeV, without any evidence of a cut-off at TeV energies. Moreover, for all these facilities, the source appears point-like because of the limited angular resolution.

SNR G0.9+0.1 is considered to be a young PWN with an estimated age of $\sim2000-3000$ years \citep{camilo2009, sidoli00}. Due to the projected position of the source, in the direction of the Galactic Center, and the uncertainties in the electron density model in that direction, the distance is not well determined \citep[between 8 and 16 kpc, as suggested by][]{camilo2009}. SNR G0.9+0.1 has been often adopted as a benchmark to test various theoretical models \citep[e.g.][]{venter&dejager2007, qiao2009, fang&zang2010, tanakatakahara2011, rensburg2018, torres2014, zhu2018}. In the early studies of \cite{venter&dejager2007} and \cite{qiao2009}, only an approximate treatment of the energy losses was included, while the dynamical evolution of the nebula was not considered. \cite{fang&zang2010} incorporated the dynamical evolution of the nebula, but assumed an injection spectrum for the electrons in the form of a Maxwellian plus a power-law tail, instead of the most widely adopted broken power law (as in \citealt{tanakatakahara2011}, \citealt{torres2014}, and \citealt{zhu2018}). More recently, \cite{rensburg2018} presented a more accurate multi-zone time-dependent leptonic model to reproduce the spatial properties of the source. In this paper, we did not focus on modelling in detail the energy-dependent morphology of SNR G0.9+0.1 (the angular resolution at VHE is not sufficient to do it), but adopted a one-zone time-dependent leptonic model, even if it has been shown that lower energy observations with a better angular resolution would benefit from multi-zone models \citep[see e.g.][]{rensburg2018, lu19, rensburg20}. Following \cite{torres2014} and \cite{zhu2018} we considered the evolution of a single population of accelerated electrons inside an expanding uniform medium in spherical symmetry. This approach turned out to be sufficiently accurate for reproducing the multiwavelength (MWL) emission of the PWN and allowed us to make predictions on the spectrum of SNR G0.9+0.1 at the highest energies.

Similarly, we used SNR G0.9+0.1 as a test case to demonstrate the improvements that the CTA South array will allow us to achieve. The source position, its faintness (only about $2\%$ of the Crab flux) and the small angular size make this object a really interesting target for testing the capabilities of the CTA. Since the extension of the PWN in SNR G0.9+0.1 is comparable to the best angular resolution achievable with CTA, we expect to be ale to measure its size at VHEs. A measurement of the angular size of the source is needed to better constrain the physical models and to compare the source size at different wavelengths. This would help understanding if the VHE emission comes from the central source or if there is some contribution from the SNR shell. In addition, the sensitivity of CTA will be much better up to and above 100 TeV \citep{scienceCTA}, allowing us to measure a possible cut-off at energies higher than 20 TeV (not excluded with the currently available data). Also this measurement is important to better constrain the physical models of the nebula, since it will constrain the particle injection spectrum, and specifically the maximum energy of the electrons (assuming a leptonic model). At such high energies, the inverse Compton emission may be in the Klein-Nishina regime, and thus obtaining such a measurement will be a very good proxy of the actual maximal electron energy. This in turn may constrain the acceleration process at the PWN termination shock.

In this work we present a comprehensive study of the spatial and spectral properties of SNR G0.9+0.1 aiming at  testing the observability of specific features in the simulated data, studying the spatial extension of the TeV emission and the presence of a VHE cut-off in the spectrum, and comparing the data to models of the MWL spectrum. 
Furthermore, we estimate the systematic uncertainties that may affect observations of SNR G0.9+0.1 carried out with CTA.

This paper is organized as follows. In Section \ref{sec:sim} we describe the models and the analysis of the spatial and spectral properties of SNR G0.9+0.1 as seen by CTA. In Section \ref{sec:sim_res} we report the results of the simulations.  In Section \ref{sec:systematics} we estimate the systematic uncertainties and in Section \ref{sec:sys_res} we discuss the results of our analysis. In Section \ref{sec:model} we describe the implementation of a physical model for the emission of a young PWN inside a SNR. Finally in Section \ref{sec:conclusion} we discuss our results and compare the numerical solutions with the simulations of the CTA observations of SNR G0.9+0.1.

\section{Simulations}
\label{sec:sim}

To simulate, reduce and analyze the $\gamma$-ray data we made use of the software ctools, a software package developed for the scientific analysis of CTA data \citep{2016A&A...593A...1K}.

We specified in input: a spatial and a spectral model describing the emission region of SNR G0.9+0.1 and a model for the spatial distribution of the cosmic-ray background.
For the spectral models we adopted both a power law and a power law with an exponential cut-off (PLEC):
 \begin{equation} 
  \frac{dN}{dE} = N_0 \biggl(\frac{E}{E_0}\biggl)^{-\Gamma},
  \label{eq:pwl}
\end{equation}   
 \begin{equation} 	  
\frac{dN}{dE} = N_0 \biggl(\frac{E}{E_0}\biggl)^{-\Gamma} \exp\biggl(\frac{E}{E_{cut}}\biggl),
\label{eq:plec}
\end{equation}   
where $N_0$ is a normalization factor, $\Gamma$ the spectral index, $E_0$ the pivot energy and $E_{cut}$ the cut-off energy.
For the spatial model, we use different distributions as described in the following.

SNR G0.9+0.1 is projected in the direction of the crowded region of the Galactic Center. In order to understand which sources can significantly affect the measurement of the flux of SNR G0.9+0.1 and to test the capability of ctools in reproducing the extended emission of the Galactic Center, we simulate the $\gamma$-ray emission in a field of \ang{3}x\ang{1} around the position of Sgr A*. In doing that, we take into account all the known TeV sources and the diffuse emission in the direction of the Galactic Center, as outlined below.

\subsection{Galactic center extended region}
\label{sec:sim_gc}

In a box of 3 square degrees around the center of the Galaxy, there are many sources at TeV energies as observed by the HESS, MAGIC and VERITAS collaborations \citep{2006Natur.439..695A, VERITAS_2016_GC, 2017arXiv170604535H, 2017A&A...601A..33A}.

We consistently selected all the sources from the HESS  catalogue\footnote{\url{www.mpi-hd.mpg.de/hfm/HESS/pages/home/sources/}}, except for SNR G0.9+0.1 for which we considered all the data included in a joint HESS$+$VERITAS\footnote{\url{veritas.sao.arizona.edu/}} analysis of the source \citep{2015arXiv150806311S}.\footnote{The results of the analysis on the sole HESS data \citep{2005A&A...432L..25A} are consistent with the results of the joint analysis.}
The sources considered in our simulation are listed below and their spatial and spectral parameters are reported in Table \ref{tab:input_par}.

\begin{itemize}
  \item{HESS J1745-290 \citep{2004A&A...425L..13A}: This source represents the TeV emission coming from the center of our Galaxy \citep{2010MNRAS.402.1877A}. It is associated with the super-massive black hole Sgr A* or to the candidate PWN G359.95-0.04 \citep{2015arXiv151101159K}. It is modelled as a point source with a power law spectrum with an exponential cut-off.
  The spectral parameters are taken from \citet{2009A&A...503..817A}}
						 
  \item{HESS J1741-302 \citep{2008AIPC.1085..249T}: It is an unidentified source detected with HESS at $\sim1\%$ of the Crab flux above 1 TeV. We modelled it as a point source with a power law spectrum.}
						    
  \item{HESS J1745-303 \citep{2006ApJ...636..777A}: This is an extended and unidentified VHE $\gamma$-ray source at a Galactic longitude of \ang{-0.4}. The morphology of the source is quite complex owing to the presence of 3 major emitting regions.
  The spatial extension of this source has been modelled using the HESS excess map\footnote{\url{www.mpi-hd.mpg.de/hfm/HESS/pages/publications/auxiliary/hessj1745-303-aux.html}}, shown in Figure \ref{fig:j1745-303}.
						    \begin{figure}
						    \centering
						    \includegraphics[width=0.9\columnwidth]{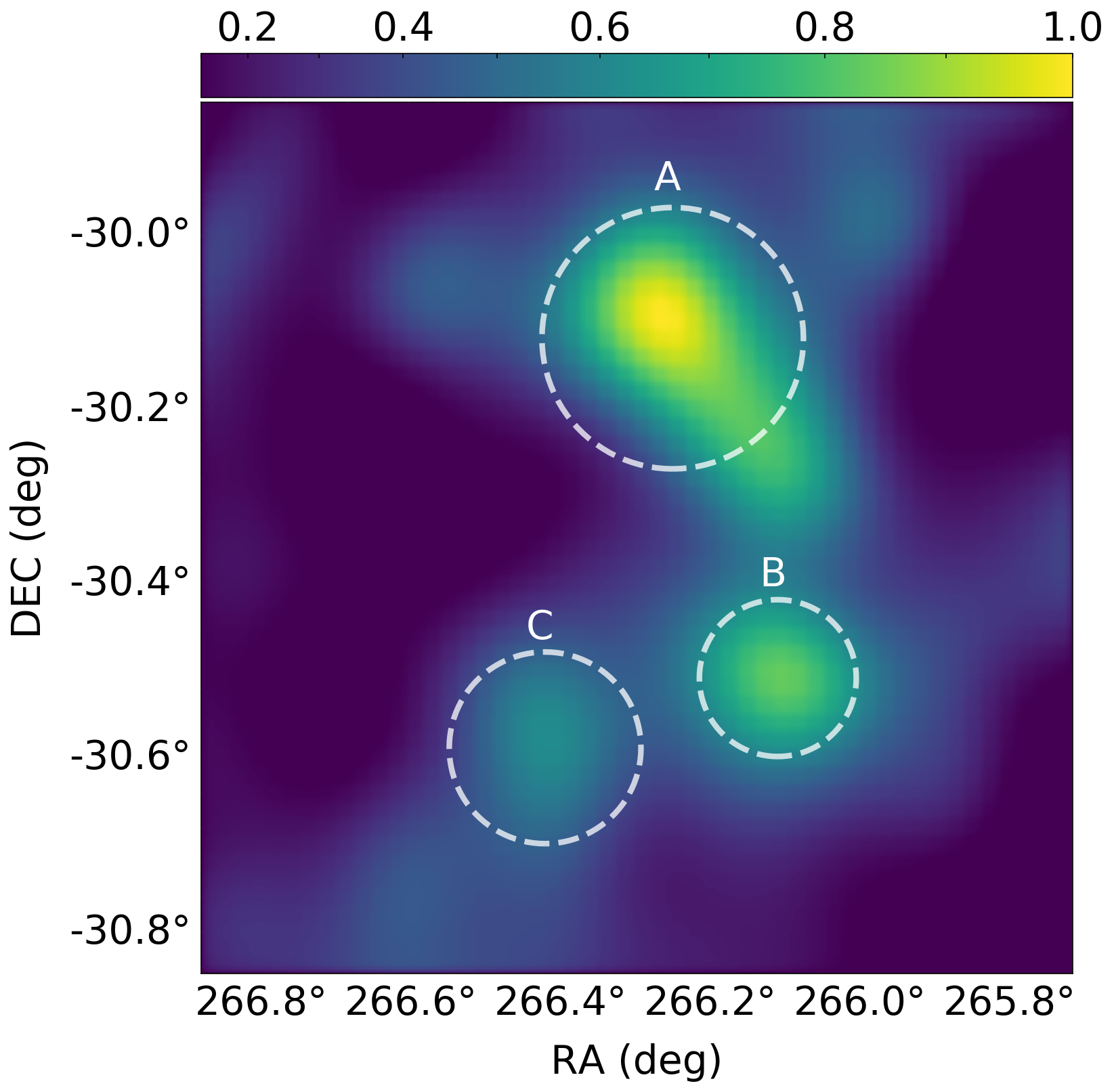}
						    \caption{Excess map of HESS J1745-303 \citep{2008A&A...483..509A}, used as spatial model for our simulations.The three dashed circles indicate the positions of the brightest emitting regions of the source.}
						    \label{fig:j1745-303}
						    \end{figure}
  The spectral model is a power law \citep{2008A&A...483..509A}.}

  \item{Diffuse emission along the Galactic plane \citep{2006Natur.439..695A}: It is a region of diffuse emission (of approximately $\pm\ang{1}$ in galactic longitude) probably associated with the interaction of 
  cosmic-ray particles with molecular clouds and that contains a number of unidentified sources such as for example HESS J1746-285 \citep{2017arXiv170604535H}}. This diffuse emission is the only source that can affect our simulation of SNR G0.9+0.1 because the spatial emission regions of these sources overlap.
  For the spatial model we used a section (between $\ang{359.1}<l<\ang{1.5}$ and $|b|<\ang{0.4}$, in Galactic coordinates) of an image taken from HESS\footnote{\url{www.mpi-hd.mpg.de/hfm/HESS/pages/publications/auxiliary/gcdiffuse_auxinfo.html}} (Figure \ref{fig:hess_res}) in which the emission coming from HESS J1745-290 and SNR G0.9+0.1 has been previously subtracted.
									     \begin{figure}
									     \centering
									     \includegraphics[width=1\columnwidth]{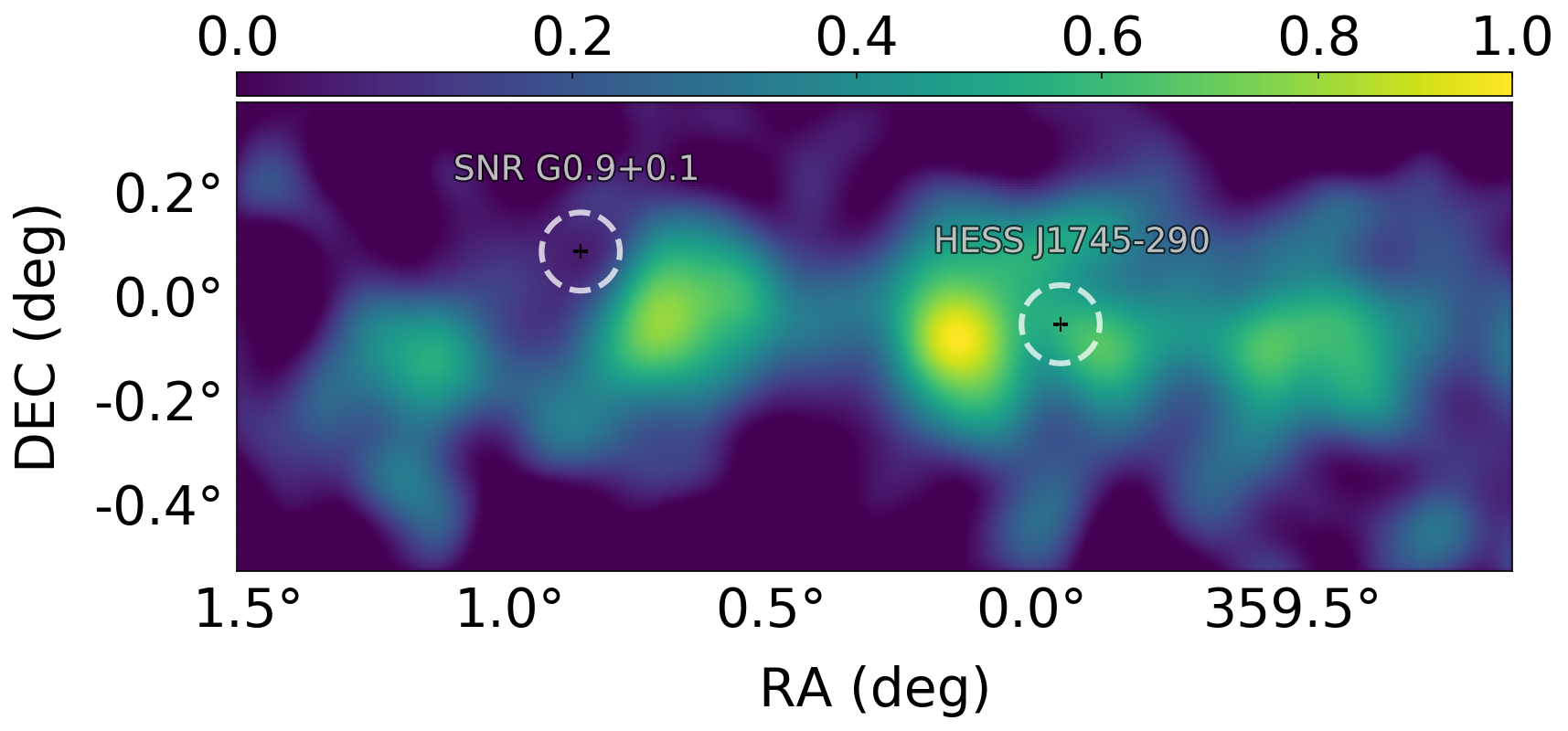}
									     \caption{HESS excess map of the diffuse emission around the Galactic center (the emission from SNR G0.9+0.1 and HESS J1745-209 has been previously subtracted) \citep{2006Natur.439..695A}, used as input spatial model for our simulations.}
									     \label{fig:hess_res}
									     \end{figure}
  The spectral model is a power law.

  \item{SNR G0.9+0.1 \citep{2005A&A...432L..25A}: The spatial model is taken from a radio map at 843 MHz from the Sydney University Molonglo Sky Survey \footnote{\url{skyview.gsfc.nasa.gov/surveys/sumss/mosaics/Galactic/J1752M28.FITS}} (radiomap template hereafter). The map has been prepared for the simulation with a technique developed for the analysis of extended sources in \textit{Fermi}-LAT\footnote{\url{fermi.gsfc.nasa.gov/ssc/data/analysis/scitools/extended/extended.html}}.
						    \begin{figure}
						    \centering
						    \includegraphics[width=0.9\columnwidth]{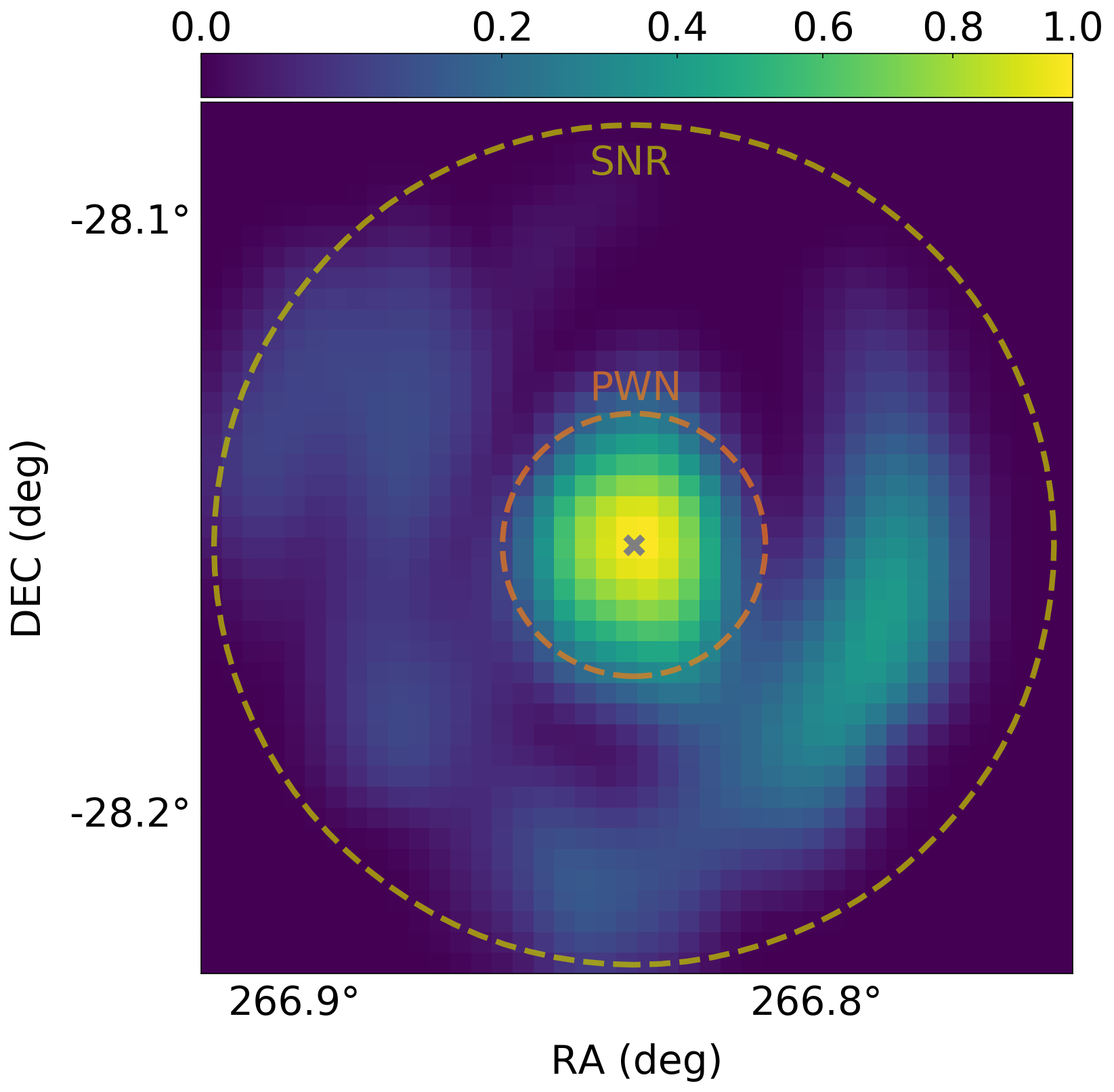}
						    \caption{Radio image (843 MHz) of SNR G0.9+0.1 taken from the Sydney University Molonglo Sky Survey (SUMSS) and used as template for some of our simulations. Most of the power in the radio band is coming from the PWN that is surrounded by the less energetic shell of the supernova remnant.}
						    \label{fig:SUMSS}
						    \end{figure}
  For the spectral model we used a single power-law for the entire system, as assumed for the HESS and VERITAS observations \citep{2015arXiv150806311S} since from currently available data it is not possible to discriminate between the emission coming from the PWN and the SNR.}
\end{itemize}

\begin{table*}
	\centering
	\caption{Input data used with \textit{ctobssim} to simulate the VHE emission from a region of $\approx\ang{3}x\ang{1}$ around the Galactic center. The reported positions are taken from the SIMBAD astronomical database 
		 \citep{2000A&AS..143....9W}, except for the position of the Galactic diffuse emission for which we adopt the position of the center of the template map. $E_0$ is always equal to 1 TeV.}
	\label{tab:input_par}
	\def\arraystretch{1.2}%
	\begin{tabular}{lcccl} 
		\hline
		\hline
		Source & Spatial Model & Position & Spectral Model & \multicolumn{1}{c}{Input Parameters} \\
		\hline
		\multirow{3}*{HESS J1745-290} & \multirow{3}*{point source}  & \multirow{2}*{RA=$266^\circ_\cdotp4150$} & \multirow{3}*{PLEC\textsuperscript{d}} & $N_0 = 2.55\times 10^{-12}$ TeV$^{-1} $cm$^{-2} $s$^{-1}$ \\
					      & 			     & \multirow{2}*{Dec=$-29^\circ_\cdotp0061$} & & $\Gamma = 2.10$ \\
					      &			           & & & $E_{cut}=15.7$ TeV\\
		\hline
		\multirow{2}*{HESS J1741-302} & \multirow{2}*{point source}  & \multirow{1}*{RA=$265^\circ_\cdotp2500$}  & \multirow{2}*{power law\textsuperscript{e}} & $N_0 = 2.34\times 10^{-13}$ TeV$^{-1}$cm$^{-2}$s$^{-1}$\\
		                        &    		                     & \multirow{1}*{Dec=$-30^\circ_\cdotp2000$} &                          & $\Gamma = 2.30$\\
		\hline
		\multirow{2}*{HESS J1745-303} & \multirow{1}*{Extended Source}  & \multirow{1}*{RA=$266^\circ_\cdotp2970$} & \multirow{2}*{power law\textsuperscript{f}} & $N_0=2.84\times10^{-12}$ TeV$^{-1}$cm$^{-2}$s$^{-1}$\\
				              & \multirow{1}*{HESS excess map\textsuperscript{a}}  & \multirow{1}*{Dec=$-30^\circ_\cdotp1990$} &                	      & $\Gamma = 2.71$ \\
		\hline
		\multirow{2}*{Galactic Diffuse} & \multirow{1}*{Extended Source}  & \multirow{1}*{RA=$266^\circ_\cdotp6518$} & \multirow{2}*{power law\textsuperscript{g}} & $N_0=1.73\times10^{-8}$ TeV$^{-1}$cm$^{-2}$s$^{-1}$sr$^{-1}$\\
						& \multirow{1}*{HESS excess map\textsuperscript{b}}  & \multirow{1}*{Dec=$-28^\circ_\cdotp7166$} &                 & $\Gamma = 2.29$\\
		\hline
 		\multirow{2}*{SNR G0.9+0.1} & \multirow{1}*{Extended Source}  		& \multirow{1}*{RA=$266^\circ_\cdotp8250$} & \multirow{2}*{power law\textsuperscript{h}} & $N_0=8.80\times10^{-13}$ TeV$^{-1}$cm$^{-2}$s$^{-1}$\\
					    & \multirow{1}*{SUMSS radio map (843 MHz)\textsuperscript{c}}  & \multirow{1}*{Dec=$-28^\circ_\cdotp1500$} &                 & $\Gamma = 2.30$ \\
		\hline
		\multicolumn{5}{l}{\scriptsize{\text{\textsuperscript{a}Figure \ref{fig:j1745-303}, \textsuperscript{b}Figure \ref{fig:hess_res}, \textsuperscript{c}Figure \ref{fig:SUMSS}}}}\\
		\multicolumn{5}{l}{\scriptsize{\text{\textsuperscript{d}\cite{2009A&A...503..817A}, \textsuperscript{e}\cite{2008AIPC.1085..249T}, \textsuperscript{f}\cite{2008A&A...483..509A}, \textsuperscript{g}\cite{2006Natur.439..695A}, 
		\textsuperscript{h}\cite{2015arXiv150806311S}}}}
	\end{tabular}
\end{table*}

To simulate observations of the field with the southern CTA facility (CTA-South) we made use of the Instrument Response Functions (IRFs) of the baseline array made available by the CTA Consortium \citep[][]{acharyya19}.
We provided as input all the information on the sources listed above, plus a model for the spatial distribution of the cosmic-ray background ("CTAIrfBackground").
We simulated four observations with different observing times centered on the position of Sgr A*, in the energy range $0.2-180$ TeV: one 30 minute observation, one 5 hour observation, one 50 hours observation and one 200 hours observation.
We simulated observations lasting up to 200 hours because we wanted to test the results achievable with CTA under the best assumptions regarding the observing time. This number is justified by the fact that the Galactic Center will be extensively observed during the first years of CTA operations \citep{scienceCTA}.
We then made an unbinned analysis\footnote{\url{http://cta.irap.omp.eu/ctools/users/tutorials/quickstart/unbinned.html}} and fitted all the simulated data with the same models given in input. Applying the maximum likelihood method, we finally compute the Test Statistics (TS) value for each source\footnote{The square root of the TS value is roughly the Gaussian $\sigma$ in the case of one free parameter associated to the source \citep[see e.g.][]{2002ApJ...571..545P}}.

\subsection{SNR G0.9+0.1}
\label{sec:sim_pwn}

As far as SNR G0.9+0.1 is concerned, we divided the analysis in two parts: first we fixed all the spectral parameters of the source and varied only the spatial model, then we kept fixed the spatial model (one of the previously selected models) and varied the spectral parameters.
At this stage we include in the simulations only the information on SNR G0.9+0.1, the diffuse emission from the Galactic plane and the cosmic-ray background.
The simulated field has a radius of \ang{0.25} centered on the source. 

To understand the capabilities of CTA in resolving the spatial extension of the VHE emission of SNR G0.9+0.1, we perform the simulations using different spatial models in the energy range 0.2-180 TeV.
All the simulated observations last 200 hours and have fixed spectral parameters (a power law with the parameters reported in Table \ref{tab:input_par}).  
The spatial models used here are: point source (assuming that the VHE emission comes only from the inner part of the remnant), a radio map template (assuming that the VHE emission comes from the same region as the radio emission) and various spatially uniform radial disk models with different radii, from 1 arcsec to 90 arcsec.
We then fit all the simulated data with four different spatial models: a point source model, a radial Gaussian model, a radial disk model, and the radiomap template model. Model fitting has been performed with a binned maximum likelihood analysis\footnote{\url{http://cta.irap.omp.eu/ctools/users/tutorials/quickstart/fitting.html}} (binned cube centered on source position with \ang{0.01} pixel size bin, 2500 pixel, gnomonic projection, and 100 logarithmic energy bins). At this stage, we adopted the binned analysis because, for long exposures, the computation time is much shorter than with the unbinned analysis.

After the analysis of the spatial properties of the source, we perform the analysis of the spectral properties fixing all the spatial parameters.
Our goal is to asses the detectability of the source in the higher energy range (from 30 TeV up to 180 TeV) and the capability of CTA-South to distinguish between different spectral models.
We simulate different observations, all lasting 200 hours, with the source spatially modelled with the radiomap template and spectrally modelled with a power law and various PLEC with different cut-off energies (20 TeV, 30 TeV, 50 TeV, and 100 TeV).
Data are simulated in the energy range between 0.2 TeV and 180 TeV. 
Model fitting has been performed with the binned likelihood analysis. The spectral energy distribution (SED) of the source is extracted using \textit{csspec}, a specific tool of ctools. 

\section{Simulations results}
\label{sec:sim_res}

\subsection{Galactic center extended region}

In Table \ref{tab:output_par} we show the results of the unbinned analysis performed on the four different simulations of the Galactic center region mentioned in section \ref{sec:sim_gc}. We report all the TS values and the spectral parameters measured for all the sources in the simulations.
These measurements were performed to check the detectability of all the simulated sources and to determine the needed observing time to reliably recover all the parameters of the sources. After 30 minutes of observation, all the sources are significantly detected and, as expected, the significance grows increasing the observing time. Already at 50 hours the inferred parameters are in good agreement with the input ones. At 200 hours, the inferred parameters are very close to the simulated ones and the associated errors become very small. Therefore, a 200 hours observation would lead to the accuracy on the measured parameters of SNR G0.9+0.1 needed for the analysis reported below.

\begin{figure}
	\centering
	\includegraphics[width=1\columnwidth]{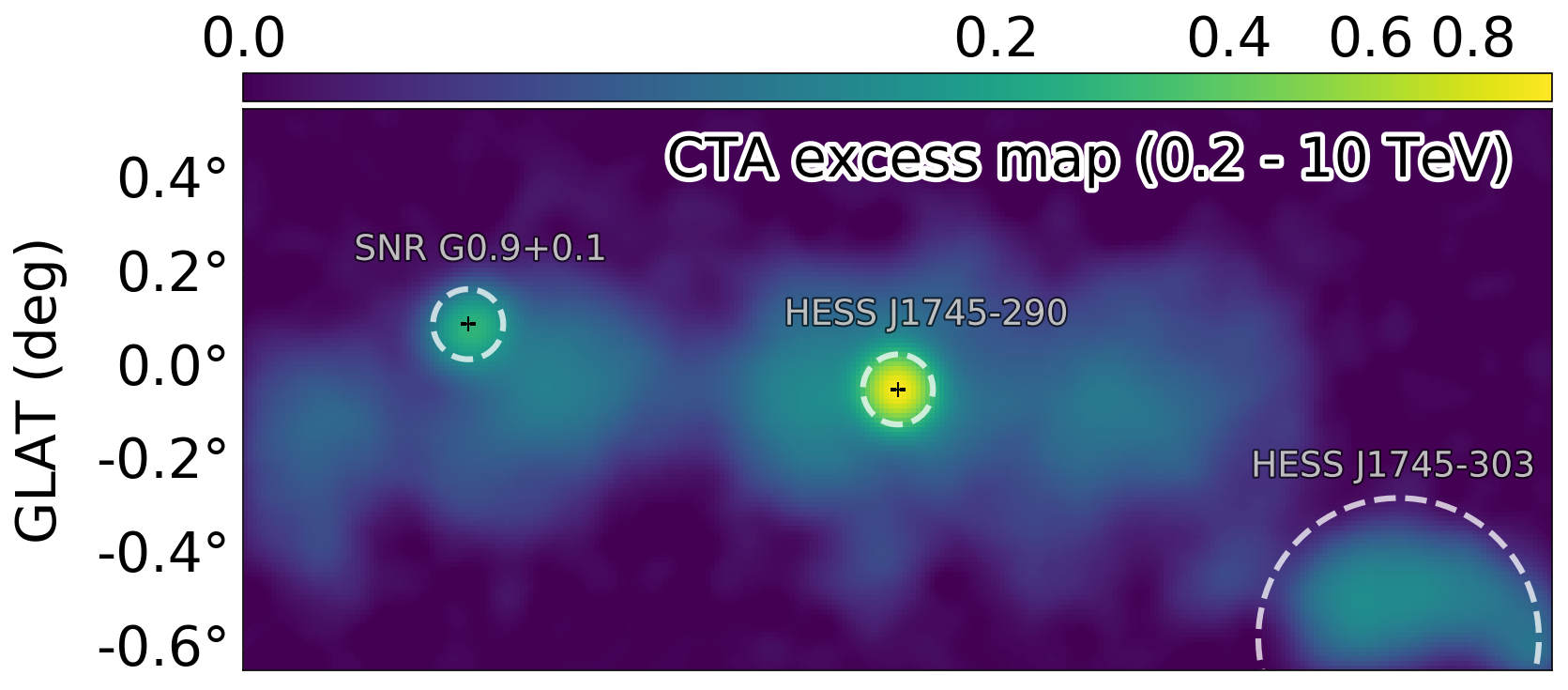}
	\includegraphics[width=1\columnwidth]{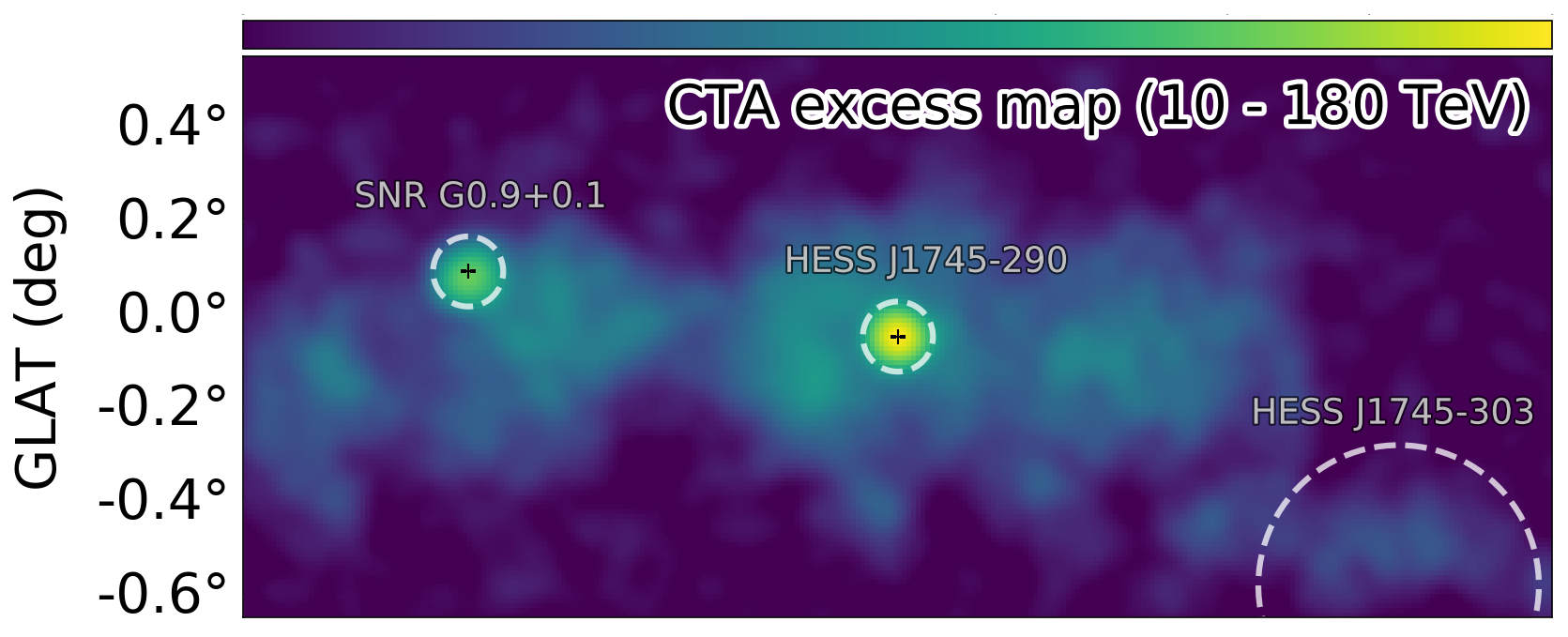}
	\includegraphics[width=1\columnwidth]{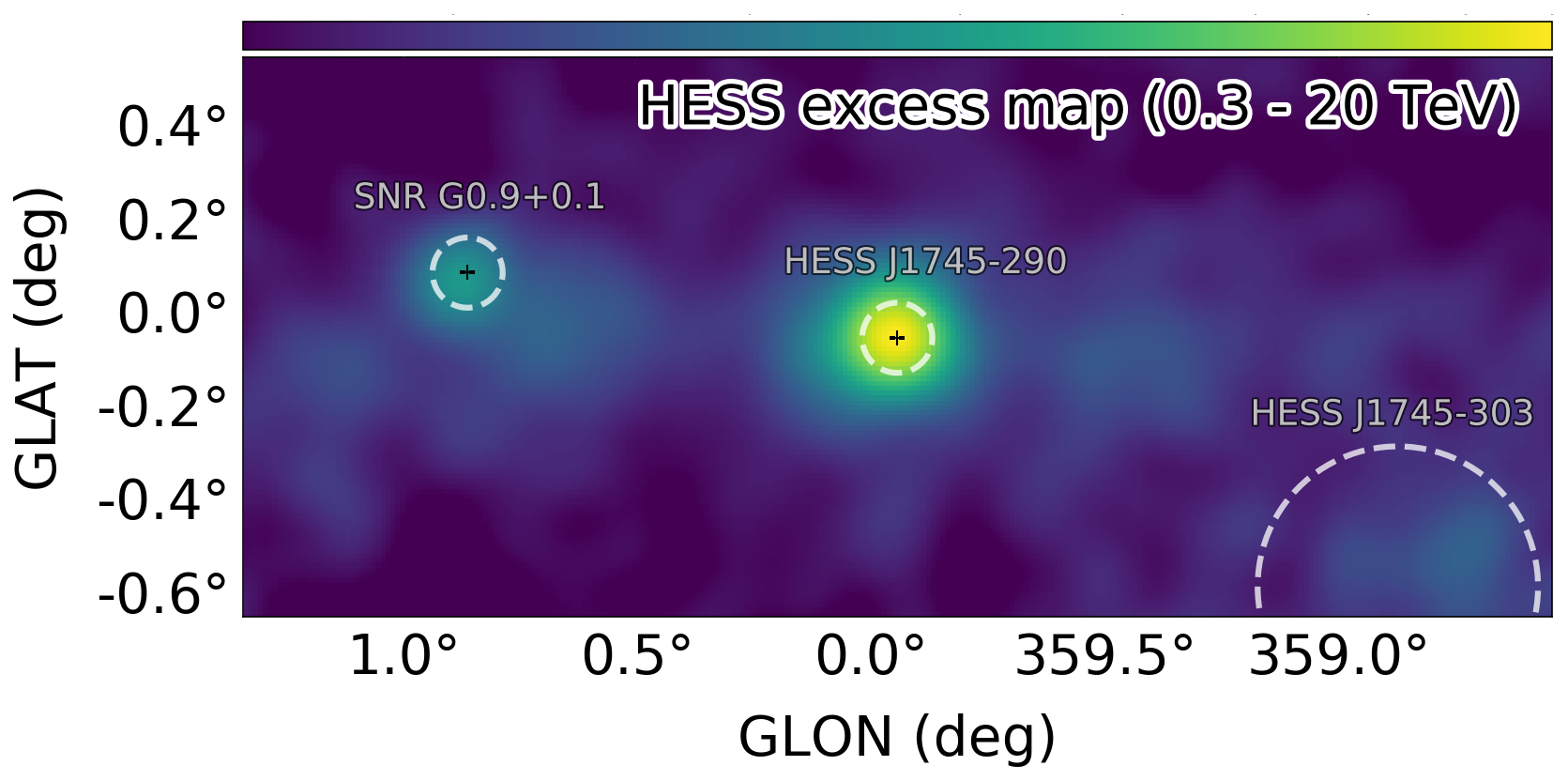}
	\caption{Simulation of the Galactic center extended emission as seen with CTA in an observation of 50 hours in two different energy ranges ($0.2-10$ TeV top panel, $10-180$ TeV middle panel) and a residual map of the same region from an HESS observation in the energy range $\sim0.3-20$ TeV \citep[lower panel; ][]{2006Natur.439..695A}}.
	\label{hessVScta}
\end{figure}

\begin{table*}
	\centering
	\caption{Results of the unbinned maximum likelihood analysis on the simulated observations of the Galactic center region. After 30 minutes of observation all the sources are significantly detected.}
	\label{tab:output_par}
	\def\arraystretch{1.2}
	\begin{tabular}{lllll}
		\hline
		\hline
		\multirow{2}*{Source} & \multicolumn{2}{c}{0.5 hour observation} & \multicolumn{2}{c}{5 hour observation} \\
				      & \multicolumn{1}{l}{Spectral parameters\textsuperscript{a,b}} & TS & \multicolumn{1}{l}{Spectral parameters\textsuperscript{a,b}} & TS \\
		\hline
		\multirow{3}*{HESS J1745-290} & $N_0=(3.29\pm0.47)\times10^{-12}$ & \multirow{3}*{371} & $N_0=(2.51\pm0.11)\times10^{-12}$ & \multirow{3}*{3444} \\
					      & $\Gamma = 1.95\pm0.12$	&			       & $\Gamma = 2.07\pm0.04$	\\
					      & $E_{cut}=5.73\pm2.06$ &			       & $E_{cut}=14.54\pm3.53$ \\
		\hline
		\multirow{2}*{HESS J1741-302} & $N_0=(2.36\pm0.74)\times10^{-13}$ & \multirow{2}*{27} & $N_0=(2.55\pm0.26)\times10^{-13}$ & \multirow{2}*{301} \\
					      & $\Gamma = 2.32\pm0.29$	&			       & $\Gamma = 2.21\pm0.07$	\\
		\hline
		\multirow{2}*{HESS J1745-303} & $N_0=(2.49\pm0.28)\times10^{-12}$ & \multirow{2}*{318} & $N_0=(2.73\pm0.09)\times10^{-12}$ & \multirow{2}*{3081} \\
					      & $\Gamma = 2.83\pm0.08$	&			       & $\Gamma = 2.71\pm0.02$ \\
		\hline
		\multirow{2}*{Galactic Diffuse} & $N_0=(1.86\pm0.06)\times10^{-8}$sr$^{-1}$ & \multirow{2}*{869} & $N_0=(1.71\pm0.01)\times10^{-8}$sr$^{-1}$  & \multirow{2}*{7787} \\
					      & $\Gamma = 2.31\pm0.03$	&			       & $\Gamma = 2.30\pm0.01$	\\
		\hline
		\multirow{2}*{SNR G0.9+0.1} & $N_0=(7.66\pm1.32)\times10^{-13}$ & \multirow{2}*{103} & $N_0=(8.12\pm0.41)\times10^{-13}$ & \multirow{2}*{995} \\
					      & $\Gamma = 2.13\pm0.11$	&			       & $\Gamma = 2.31\pm0.04$ \\
		\hline
		\hline
                \multirow{2}*{} & \multicolumn{2}{c}{50 hour observation} & \multicolumn{2}{c}{200 hour observation} \\
				      & \multicolumn{1}{l}{Spectral parameters\textsuperscript{a,b}} & TS & \multicolumn{1}{l}{Spectral parameters\textsuperscript{a,b}} & TS \\
		\hline
		\multirow{3}*{HESS J1745-290} & $N_0=(2.61\pm0.03)\times10^{-12}$ & \multirow{3}*{36493} & $N_0=(2.53\pm0.01)\times10^{-12}$ & \multirow{3}*{142506} \\
					      & $\Gamma = 2.07\pm0.01$	&			       & $\Gamma = 2.10\pm0.01$	\\
					      & $E_{cut}=13.75\pm0.86$  &			       & $E_{cut}=15.83\pm0.53$ \\
		\hline
		\multirow{2}*{HESS J1741-302} & $N_0=(2.36\pm0.08)\times10^{-13}$ & \multirow{2}*{2526} & $N_0=(2.38\pm0.04)\times10^{-13}$ & \multirow{2}*{9987} \\
					      & $\Gamma = 2.27\pm0.03$	&			       & $\Gamma = 2.30\pm0.01$ \\
		\hline
		\multirow{2}*{HESS J1745-303} & $N_0=(2.79\pm0.03)\times10^{-12}$ & \multirow{2}*{32853} & $N_0=(2.83\pm0.01)\times10^{-12}$ & \multirow{2}*{132875} \\
					      & $\Gamma = 2.72\pm0.07$	&			       & $\Gamma = 2.71\pm0.01$ \\
		\hline
		\multirow{2}*{Galactic Diffuse} & $N_0=(1.73\pm0.01)\times10^{-8}$sr$^{-1}$ & \multirow{2}*{80674} & $N_0=(1.73\pm0.01)\times10^{-8}$sr$^{-1}$  & \multirow{2}*{322679} \\
					      & $\Gamma = 2.28\pm0.03$	&			       & $\Gamma = 2.29\pm0.01$ \\
		\hline
		\multirow{2}*{SNR G0.9+0.1} & $N_0=(8.87\pm0.13)\times10^{-13}$ & \multirow{2}*{11274} & $N_0=(8.83\pm0.06)\times10^{-13}$ & \multirow{2}*{44901} \\
					      & $\Gamma = 2.30\pm0.01$	&			       & $\Gamma = 2.30\pm0.01$ \\
		\hline
		\multicolumn{5}{l}{\scriptsize{\text{\textsuperscript{a} $N_0$ in unit of TeV$^{-1}$cm$^{-2}$s$^{-1}$ and $E_{cut}$ in unit of TeV. $E_0=1$TeV}}}\\
		\multicolumn{5}{l}{\scriptsize{\text{\textsuperscript{b} Statistical error only}}}\\
	\end{tabular}
\end{table*}
We then compared the simulation obtained for an exposure of 50 hours with that obtained with HESS in 55 hours\footnote{\url{www.mpi-hd.mpg.de/hfm/HESS/pages/publications/auxiliary/gcdiffuse_auxinfo.html}} \citep{2006Natur.439..695A} in a similar energy range (see Figure \ref{hessVScta}).
The images are in good agreement, with the CTA simulated one having a lower background contamination.
With the same observing time, CTA will allow us to obtain a wider spectral coverage and a higher signal-to-noise ratio.

\subsection{SNR G0.9+0.1}
\label{sec:simSNR}
\begin{figure}
	\centering
	\includegraphics[width=1\columnwidth]{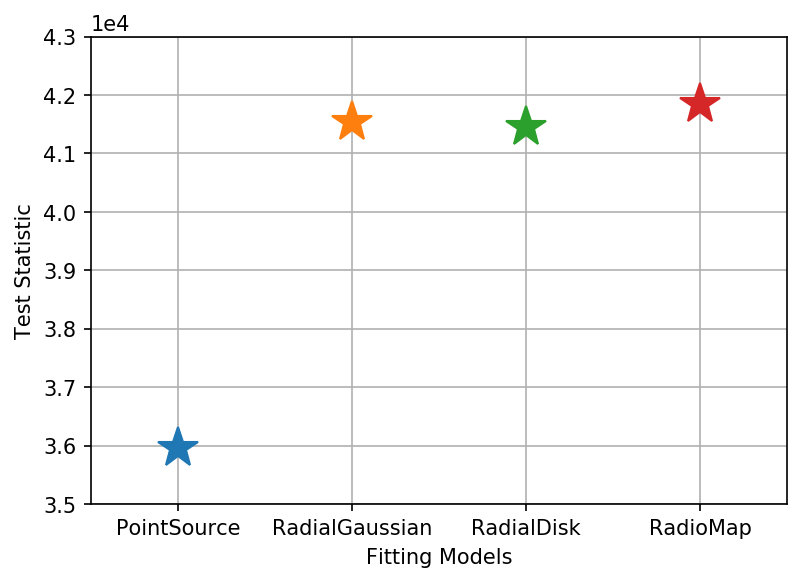}
	\caption{Test Statistics values for different fitting models applied to the simulation in which the VHE emitting region of SNR G0.9+0.1 is modelled with the radio map template. The TS for the point source fitting model has a lower significance compared to the other fitting models.}
	\label{TSradiomap}
\end{figure}

We performed two different analyses to investigate the resolving capabilities of CTA. In the first analysis, we carry out different fits of the  image simulated using the radiomap template. The fits were performed with four different spatial models: point source model, spatially uniform radial disk model (with the radius left free during the fit), radial Gaussian model (with width left free during the fit) and the radiomap model. The results are shown in Figure \ref{TSradiomap}. If the VHE emission follows the radio emission CTA could be able to detect the source as an extended object because the TS value for the point source fit is significantly lower. The extended models have similar TS values, with the radiomap template being slightly more significant, indicating that all the three models can reproduce well the simulated data and that the VHE $\gamma$-ray emission from outside the PWN (i.e. the emission coming from the SNR shell that can be seen in Figure \ref{fig:SUMSS}) is almost negligible.

\begin{figure}
	\centering
	\includegraphics[width=1\columnwidth]{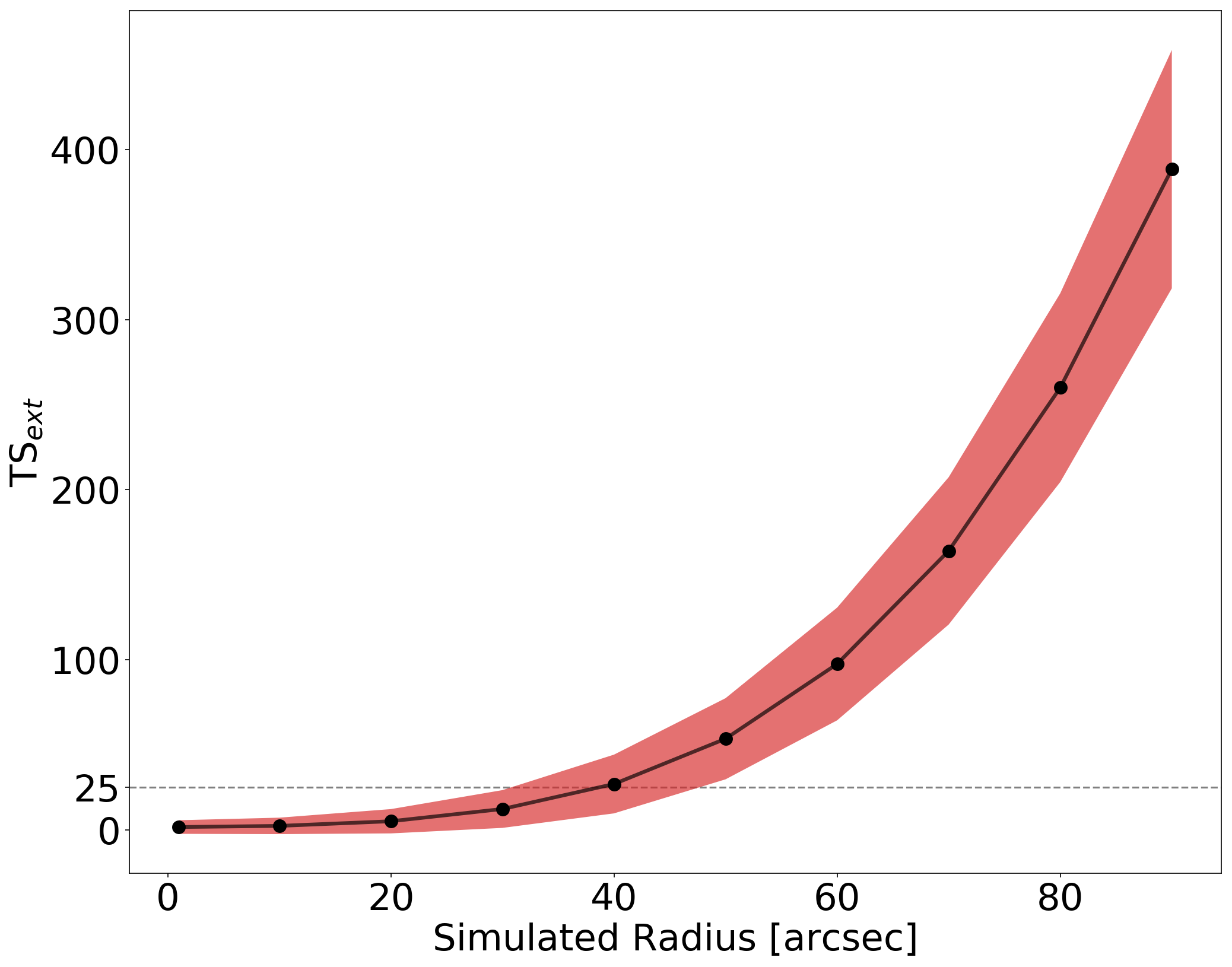}
	\caption{Significance of the detection of the source as an extended source for the images simulated using a spatially uniform radial disc with different radii. We consider $\text{TS}_\text{ext} \geq 25$ as the minimum value for claiming that the source is extended. The radial disc model has a significantly better likelihood when the source has a radius larger than $39^{+9}_{-8}$ arcsec. The red area shows the 95\% confidence level error region.}
	\label{TSradialdisk}
\end{figure}

In the second analysis, we test the limiting resolving capabilities of CTA against the background of the Galactic Center VHE emission region following the procedure developed to detect an extended source in the \textit{Fermi}-LAT data \citep{lande12}.
We have simulated different images assuming a spatially uniform radial disc with different radii. We then fit all the images with a point source model and a radial disc model with the radius free to vary. This procedure is then repeated for 100 times to account for the statistical fluctuations that can arise from different simulations\footnote{Different simulations are based on a different random seed for the Monte Carlo generator that samples the input source models to produce observed photon energies and arrival directions. This is achieved through the random number generator provided in the GammaLib library\citep{2016A&A...593A...1K}.}.
For all the simulated images we compute the significance of detecting significant spatial extension for the source by using the likelihood ratio test:
 \begin{equation} 
    \text{TS}_\text{ext} = -2\log\frac{\mathcal{L}_\text{RD}}{\mathcal{L}_\text{PS}},
\end{equation}   
where $\mathcal{L}_\text{RD}$ and $\mathcal{L}_\text{PS}$ are the likelihood values of the fits with the radial disc (RD) and the point source (PS) models. In Figure \ref{TSradialdisk} we show $\text{TS}_\text{ext}$ in function of the simulated source radius with the $95\%$ confidence level errors. The value increases from very small to large radii, showing that the radial disc model has a significantly better likelihood ($\text{TS}_\text{ext} \geq 25$) when the source has a radius larger than $39^{+9}_{-8}$ arcsec.
This means that if the VHE emission region of SNR G0.9+0.1 is bigger than $\sim0.65$ arcmin, and the response of the instrument is very well known, CTA will be able to detect it as an extended source even if the PSF of the instrument is larger ($\sim 1.8$ arcmin). 
However, it would be difficult to study substructures inside the source because the angular size of these substructures would be too small. 

\begin{figure}
	\centering
	\includegraphics[width=1\columnwidth]{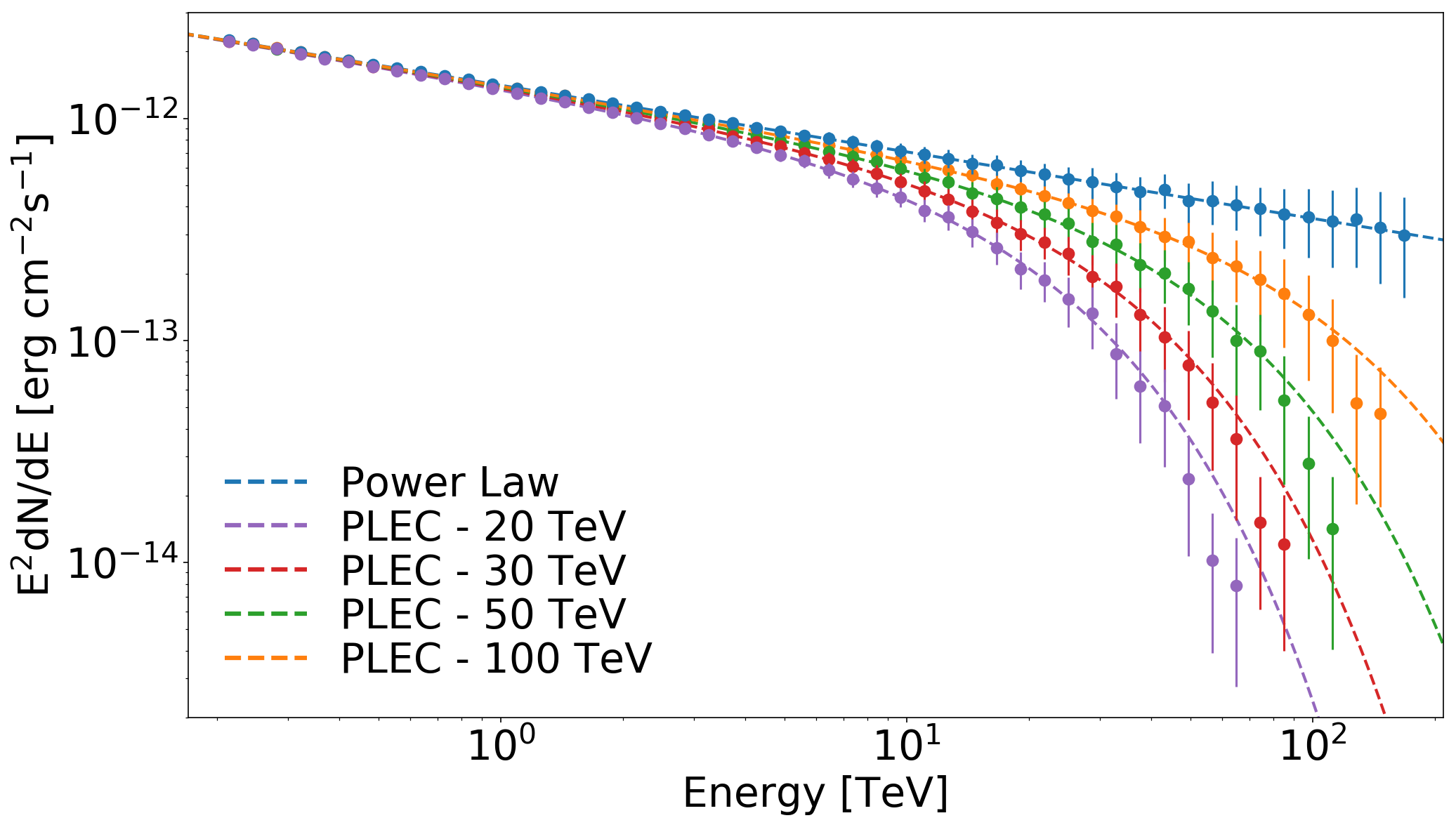}
	\caption{CTA spectral energy distribution of SNR G0.9+0.1, simulated using different cut-off energies. It is clearly possible to distinguish the different spectra.}
	\label{spectrapwl}
\end{figure}

\begin{figure}
	\centering
	\includegraphics[width=1\columnwidth]{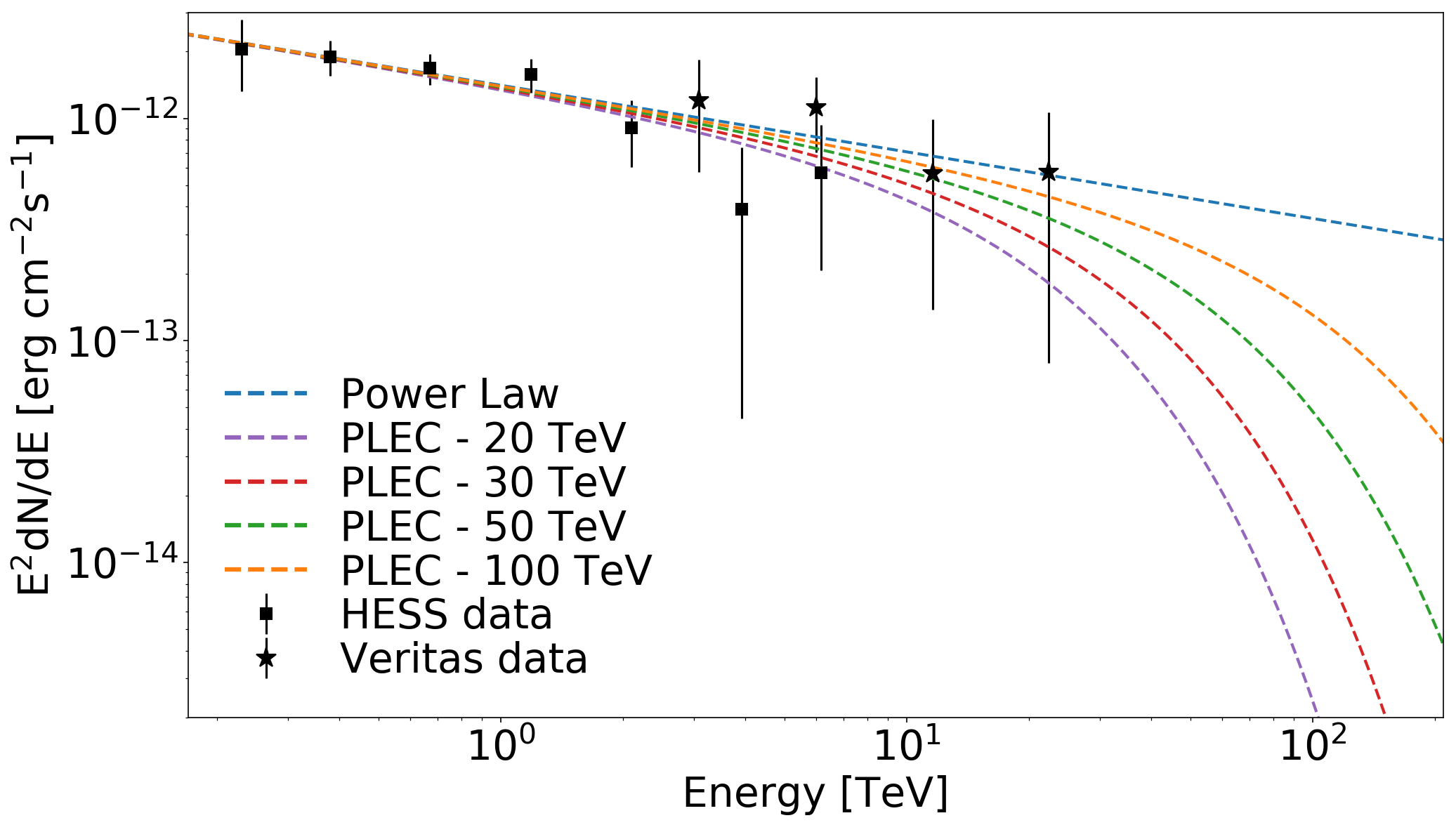}
	\caption{Comparison of the spectra simulated in this work with the data from HESS (black square) and VERITAS (black stars). With the current available data it is not possible to rule out models with cut-off energies higher than 20 TeV.}
	\label{spectraCTAvsHESSVER}
\end{figure}

As far as the CTA spectrum of SNR G0.9+0.1 is concerned, it is shown in Figure \ref{spectrapwl}.
The present analysis aims at understanding how well it is possible to recover the expected cut-off of this source. This has strong implications for the physical modeling implemented in section \ref{sec:model} since a different cut-off energy could lead to inferring different physical parameters for the nebula.
The spectrum has a good statistics and therefore the spectral resolution is very good. It is clearly possible to distinguish spectra with different cut-off energies. 
This represents a significant improvement in comparison with currently available data that does not allow to distinguish if the spectral shape of the VHE emission is a power law or a power law with a cut-off at energies higher than 20 TeV (Figure \ref{spectraCTAvsHESSVER}).

The maximum cut-off energy detectable in the CTA simulated spectrum is $\gtrsim 100$ TeV while, for the lowest energy cut-off considered here (20 TeV), the source is detectable only up to $\sim 60$ TeV.

\section{Assessing systematic errors}
\label{sec:systematics}

The spectral analysis of the simulated data returns only the statistical errors, computed from the covariance matrix of the maximum likelihood fitting procedure. But systematic errors need to be carefully accounted for in order to assess the accuracy of the results. A fit of simulated data without considering the systematic errors will lead to overestimating the goodness of the fit and to results that may not be realistic.

We considered both the instrumental sources of uncertainties and the background related uncertainties.
The instrumental sources of uncertainties are due to the imperfect knowledge of the effective area and the accuracy of the reconstructed energy scale, while the background sources of uncertainties are due to the cosmic-rays and the Galactic diffuse emission.
In order to translate uncertainties into systematic errors on fluxes and spectral indices we will make some assumptions on how these uncertainties propagates. 

In the case of the instrumental uncertainties we start from CTA technical requirements and we apply the following procedure to measure the associated errors.
\begin{itemize}
 \item \textbf{Knowledge of the effective area}.\\ Uncertainty on the effective area of the system must be $<5\%$ (from the CTA technical requirements). To estimate the effect of such an uncertainty we followed the method used by the Fermi-LAT team \citep[Sec. 5.7]{2012ApJS..203....4A}. We generate perturbed IRFs that represent the worst scenario, extract the spectral parameters and compare them to those obtained with the unperturbed IRF.
 The perturbed effective area $A'_{eff}$ is written as:
  \begin{equation} 
 A'_{eff}(E, \theta) = A_{eff}(E, \theta) \cdot (1+\xi_{_{A_{eff}}} \cdot B(E)),
 \end{equation}   
 where $A_{eff}$ is the unperturbed effective area, $\xi_{_{A_{eff}}} = 0.05$ is the uncertainty and $B(E)$ a function of the energy (bracketing function).
 Different form for $B(E)$ are adopted depending on the spectral parameter considered.
 For a simple power law, to maximize the effect on the normalization, the function $B(E)$ is written as:
  \begin{equation}\label{eq:const}
  B(E) = \pm 1,
 \end{equation}   
 while, to maximize the effects on the spectral index, the following expression is used:
  \begin{equation} 
 \label{eq:tanh}
  B(E) = \pm \tanh\left(\frac{1}{0.13}\log\left(\frac{E}{E_0}\right)\right),
 \end{equation}   
 where $E_0$ is the same pivot energy used in Equation \ref{eq:pwl} and \ref{eq:plec}.
 With these two modified IRFs we have reanalyzed the data and estimated the errors on the spectral parameters from the values obtained in the two cases.
 
 \item \textbf{Accuracy of the Energy Scale.}\\ The uncertainty on the energy of a photon event candidate must be $<6\%$ (from the CTA technical requirements). In order to estimate the errors on the spectral parameters induced by this uncertainty\footnote{In this work we have not taken into account the energy dispersion since it was computationally too expensive.}, we took the simulated data and perturbed all the photon energies as:
  \begin{equation} 
 E' = E	\cdot (1 \pm \xi_{_{E_{scale}}}),
 \end{equation}   
 where $\xi_{_{E_{scale}}}=0.06$.
 We have then analyzed these data and estimated the errors on the spectral parameters.
 \end{itemize}

\begin{figure}
	\centering
	\includegraphics[width=1\columnwidth]{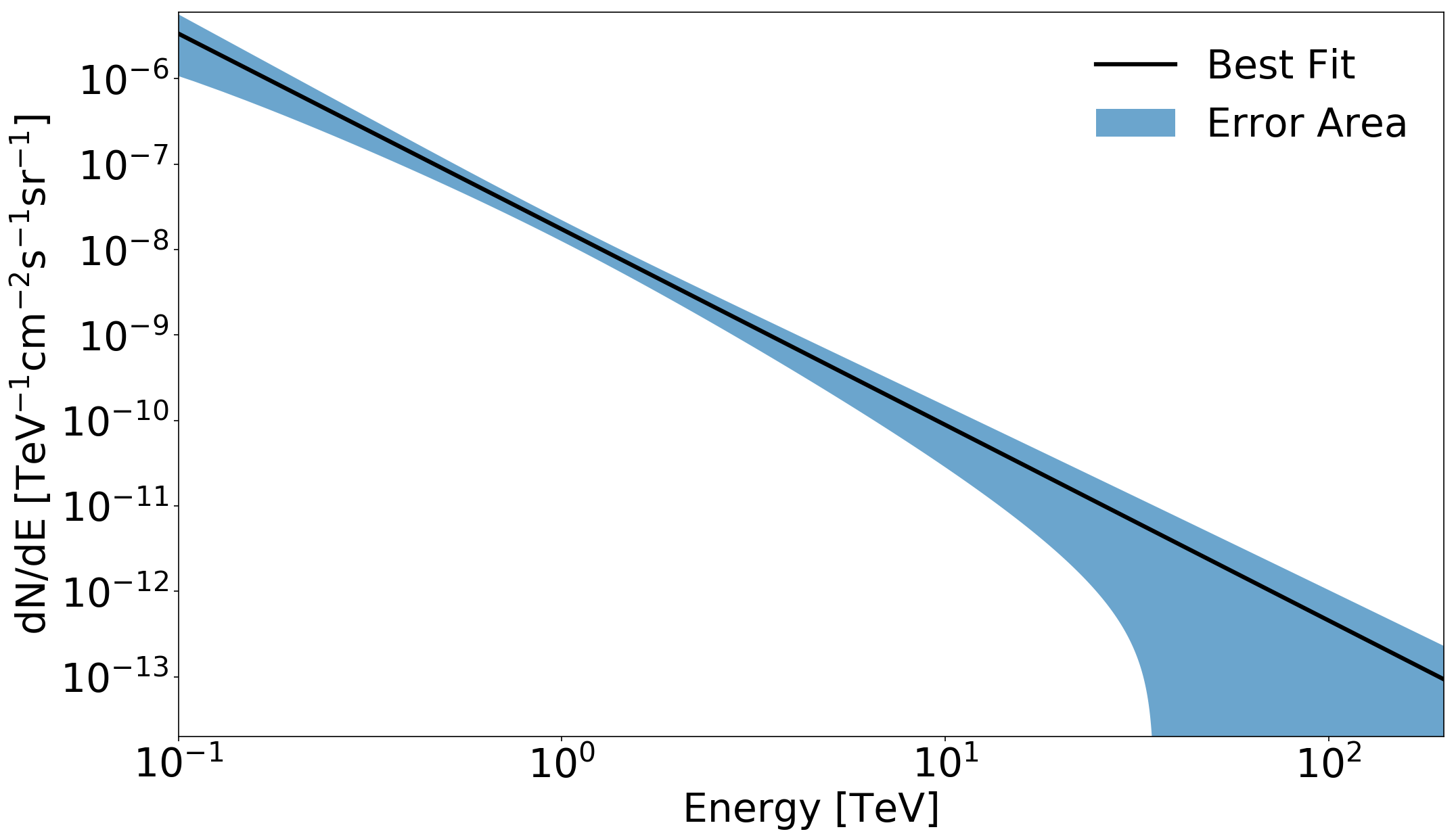}
	\caption{Average spectrum (per steradian) of the Galactic plane near the center region (between $\ang{359.2}<l<\ang{0.8}$ and $|b|<\ang{0.3}$) as measured by the HESS collaboration \citep{2006Natur.439..695A}. The shaded area correspond to the error boundary of the HESS measurements, in which both statistical and systematic errors are taken in account.}
	\label{fig:fluxGCerr}
\end{figure}

In the case of the uncertainties related to the knowledge of the background  we applied a different approach, as described below.
\begin{itemize}
\item \textbf{Cosmic-ray Background.}\\ In order to determinate the impact of the uncertainty on the cosmic-ray background we varied its flux of $\pm 50\%$\footnote{This value is much bigger then the expected uncertainty on the residual cosmic-ray background for CTA-South.} from the nominal value. We thus changed the normalization of the background according to:
  \begin{equation} 
  N'_0 = N_0\cdot(1\pm\xi_{_{CR_{bkg}}}),
 \end{equation}   
 where $\xi_{_{CR_{bkg}}} = 0.5$. We then analyzed these data and estimated the errors on the spectral parameters. Since the deviations from the nominal values resulting from this source of uncertainty seem to be negligible, as discussed in the next section, it was not worth considering variations induced by changes in the photon index of the cosmic-ray background.

 \item \textbf{Galactic Diffuse Emission.}\\ As mentioned above we modelled the emission from the Galactic plane central region using an HESS observation \citep{2006Natur.439..695A}. The best fit spectral model for this observation is a power law with $N_0=1.73\times10^{-8}$ TeV$^{-1}$cm$^{-2}$s$^{-1}$sr$^{-1}$ and $\Gamma = 2.29$ with the associated errors $\sigma_{_{N_0}} = (\pm0.13_{stat}\pm0.35_{syst})\times10^{-8}$ TeV$^{-1}$cm$^{-2}$s$^{-1}$sr$^{-1}$ and $\sigma_{_\Gamma} = \pm0.07_{stat}\pm0.20_{syst}$.
 Using these errors we calculate an optimistic/pessimistic spectrum from the Galactic center from:
  \begin{equation} 
  F_{pess,opt} = F(E) \pm \sqrt{\left(\frac{\partial F}{\partial N_0}\right)^2\sigma_{_{N_0}}^2+\left(\frac{\partial F}{\partial \Gamma}\right)^2\sigma_{_\Gamma}^2},
 \end{equation}   
 where $F(E)$ is the best fit value of the flux, the pessimistic case $F_{pess}$ corresponds to the sign $+$ and the optimistic case $F_{opt}$ to the sign $-$.
 This is an approximation of the error propagation formula (we lack all the information on the full
 covariance matrix that comes from the analysis made by the HESS collaboration).
 The spectrum is shown in Figure \ref{fig:fluxGCerr}. 
 We have then analyzed these perturbed data and measured the associated errors.
 We repeated the analysis using the pessimistic and optimistic estimate of the spectrum and used the spectral parameters of the source inferred in the two cases to estimate the errors induced by this systematic uncertainty on it. 
 It is worth to mention that also the uncertainty on the morphology of the Galactic diffuse emission can be a source of systematics error. However, at present we have not enough information to assess the uncertainties related to the
 morphology of diffuse emission. This task is left for future studies.
\end{itemize}

For all these sources of uncertainty we have repeated the simulations one hundred times and we have then taken the final errors on the average values as representative of the uncertainties induced by the different simulations.

\section{Systematic error estimation results}
\label{sec:sys_res}

In Table \ref{tab:syst_err} we report the values of the systematic errors, computed from the difference between the "Nominal value" (values computed without perturbing the data) and the values obtained as explained in the previous section. 

\begin{figure*}
	\centering
	\includegraphics[width=1\columnwidth]{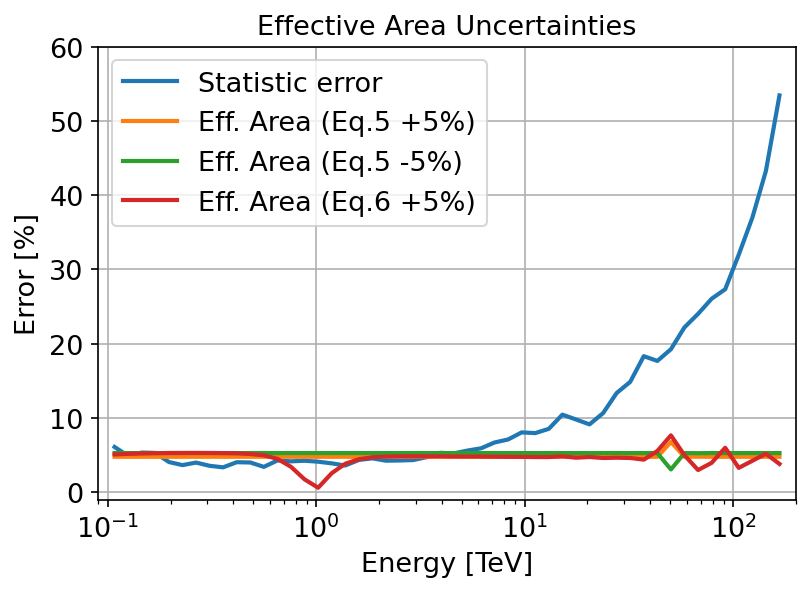}
	\includegraphics[width=1\columnwidth]{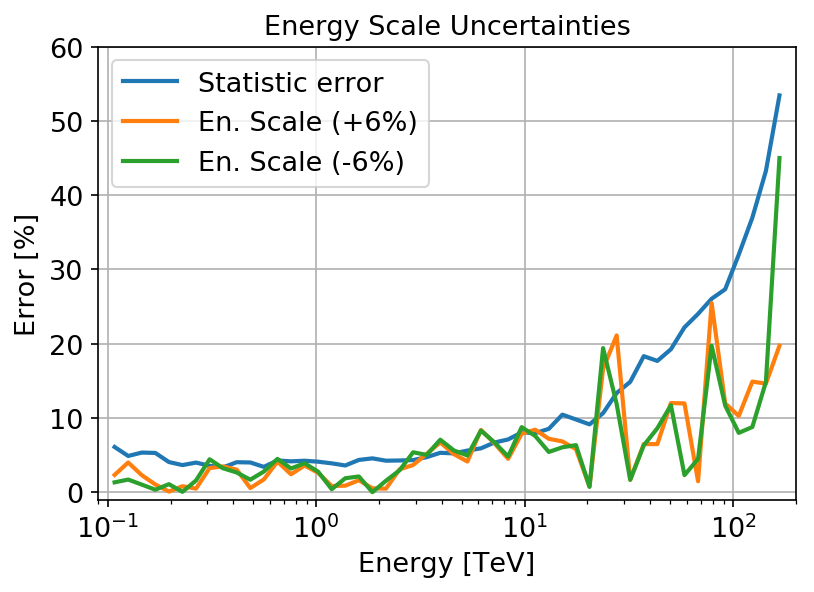}
	\includegraphics[width=1\columnwidth]{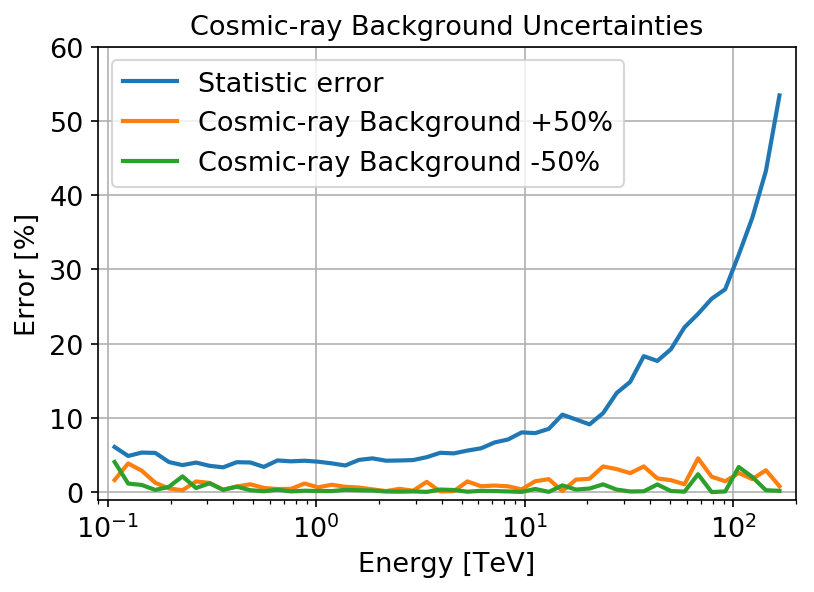}
	\includegraphics[width=1\columnwidth]{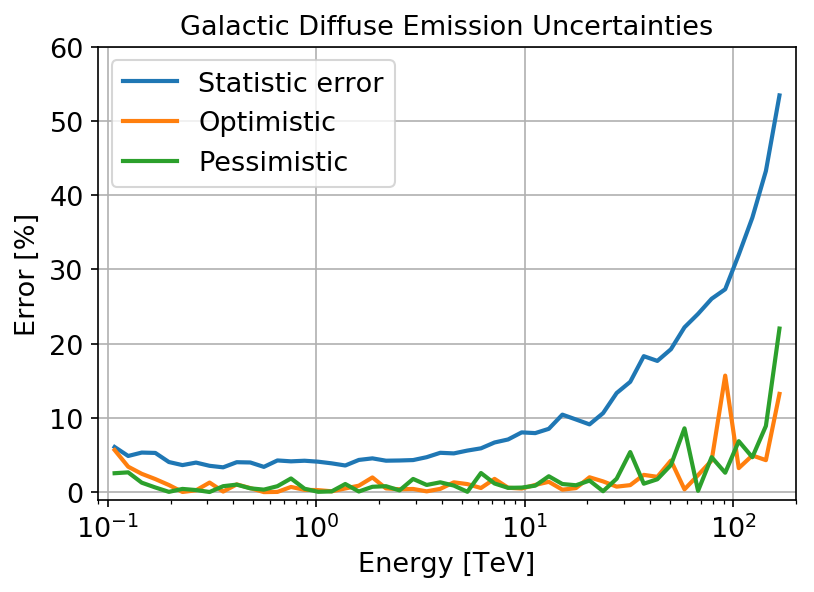}
	\caption{Fractional error on the CTA spectrum as function of photon energy, measured assuming a power law model for SNR G0.9+0.1. In the low energy range (where CTA-South will have the best sensitivity) the instrumental uncertainties are dominant, while in the higher energy range the decrease of the sensitivity of the array leads to an increase of the statistical errors. The background related uncertainties are very low at all energies.}
	\label{fig:syst_err}
\end{figure*}

The instrumental systematic uncertainties dominate over the background related sources of error. This is shown in Figure \ref{fig:syst_err} where we plot the errors as a function of energy, assuming a power law spectrum. While the systematics act differently at different energies, the background related uncertainties are always small. In the low energy range (where the array has the best sensitivity) the instrumental uncertainties dominate and are at the same level as the statistical errors, while in the higher energy range, the decrease of the sensitivity of CTA-South leads to an increase of the statistical errors.
The behavior of the statistical error yields a good representation of the sensitivity limit of the CTA-South array.

Although the errors reported here are probably overestimated (especially the instrumental ones), this analysis provides a good clue on the order of magnitude of the expected systematic uncertainties. According to the results of our analysis, the background related uncertainties are negligible in comparison with the other sources of uncertainty and have a small impact on the measured spectrum.

\begin{table*}
	\centering
	\caption{Systematic errors measured using the deviation of the perturbed values from the nominal ones, as explain in the text. We report for comparison also the statistical errors computed from the likelihood analysis made with ctools.}
	\label{tab:syst_err}
    \def\arraystretch{1.5}
	\begin{tabular}{lcccccc}
		\hline
		\hline
		\textbf{Statistical errors} & $N_0$\textsuperscript{a} & $\delta N_0$\textsuperscript{a} & $\delta N_0/N_0$ & $\Gamma$ & $\delta\Gamma$ & $\delta\Gamma/\Gamma$ \\
		\hline
		Nominal value & $8.820\times10^{-13}$ & $7.322\times 10^{-15}$ & $0.830\%$ & $2.306$ & $0.006$ & $0.251\%$\\
		\hline
		\hline
		\textbf{Systematic errors} & $N'_0$\textsuperscript{a} & $\delta N'_0$\textsuperscript{a} & $\delta N'_0/N_0$ & $\Gamma'$ & $\delta\Gamma'$ & $\delta\Gamma'/\Gamma$ \\
		\hline
		$A'_{eff}$ (Eq. \ref{eq:const} -5\%)    & $9.351\times10^{-13}$ & $5.309\times 10^{-14}$  & $6.019\%$  & $2.309$ & $0.003$  & $0.123\%$  \\
		$A'_{eff}$ (Eq. \ref{eq:const} +5\%)    & $8.334\times10^{-13}$ & $-4.865\times 10^{-14}$ & $-5.516\%$ & $2.310$ & $0.004$  & $0.177\%$  \\
		$A'_{eff}$ (Eq. \ref{eq:tanh} +5\%) & $8.846\times10^{-13}$ & $2.632\times 10^{-15}$  & $0.298\%$   & $2.333$ & $0.027$  & $1.179\%$  \\
		En. scale ($-6\%$)  & $8.555\times10^{-13}$ & $-2.656\times 10^{-14}$ & $-3.012\%$  & $2.320$ & $0.014$  & $0.613\%$  \\
		En. scale ($+6\%$)  & $9.055\times10^{-13}$ & $2.351\times 10^{-14}$  & $2.666\%$   & $2.303$ & $-0.003$ & $-0.128\%$ \\
		Cosmic-ray ($-50\%$)   & $8.837\times10^{-13}$ & $1.668\times 10^{-15}$ & $0.189\%$  & $2.302$ & $-0.042$ & $-0.079\%$ \\
		Cosmic-ray ($+50\%$)   & $8.882\times10^{-13}$ & $6.184\times 10^{-15}$  & $0.701\%$   & $2.309$ & $0.003$ & $0.117\%$ \\
		Gal. Diffuse (Opt.)        & $8.807\times10^{-13}$ & $-1.277\times 10^{-15}$  & $-0.145\%$   & $2.302$ & $-0.004$ & $-0.166\%$ \\
		Gal. Diffuse (Pess.)       & $8.807\times10^{-13}$ & $-1.315\times 10^{-15}$ & $-0.149\%$  & $2.308$ & $0.002$  & $0.082\%$  \\
		\hline
		\multicolumn{7}{l}{\scriptsize{\text{\textsuperscript{a}$N_0$ and $\delta N_0$ in unit of TeV$^{-1}$cm$^{-2}$s$^{-1}$ }}}\\
	\end{tabular}
\end{table*}

\section{Modelling the emission of Pulsar wind nebulae}
\label{sec:model}

PWNe are important laboratories to test the processes responsible for the acceleration of charged particles. To this end, it is crucial to compare real or simulated data with precise and physically motivated models.

Reproducing the broad-band spectrum, from the radio band up to $\gamma$-rays, of these sources, requires a dynamical model that describes the evolution of the population of the emitting electrons inside the PWN.

A one-zone time-dependent leptonic model is often adopted. In this model the main emitting particles are a population of electrons that evolves with time and the nebula is approximated as a sphere where the electrons are uniformly distributed.

This approach has been developed by several authors \citep[e.g.][]{venter&dejager2007, qiao2009, fang2008, fang&zang2010, tanakatakahara2010, tanakatakahara2011, bucciantini2011, martin2012, martin2016, torres2014, rensburg2018}. In this work, we follow the approach presented by \cite{gelfand2009}. We also test the result of our implementation for the PWN in SNR G0.9+0.1 against those obtained by \cite{zhu2018} and \cite{torres2014}.

\subsection{The model}
\label{sec:model_detail}

The distribution and the evolution of the electronic population inside the nebula is described by an energy-diffusion equation. The general form of this equation (see equation \ref{eq:transp_general}) and the meaning of all the terms of the equation are described in Appendix \ref{appendix:0}. The simplified form used in this work is as follow:

 \begin{equation} 
\begin{split}
    \frac{\partial N(E,t)}{\partial t} & = Q(E,t) - \frac{\partial}{\partial E}[b(E) N(E,t)] - \frac{N(E,t)}{\tau_{esc}(E,t)}.
\end{split}
\label{eq:transport}
\end{equation}   

where $N(E,t)$ is the number density of the electrons, $Q(E,t)$ the injection rate of electrons at the termination shock, $b(E)$ the variation of the mean energy of the electrons per unit time, and $\tau_{esc}(E,t)$ is a characteristic time scale describing the escape of the electrons from the system.

The typical shape adopted for the injection spectrum of the particles is a broken power law.
Other types of injection spectra have been proposed but all somehow fail to reproduce the observed spectrum or are difficult to motivate \citep[see the discussion in][]{Gelfand2017}. The broken power law spectrum can reproduce well the different slopes of the synchrotron spectrum observed in many PWNe, as the Crab Nebula \citep{atoyan&aharonian1996}, in the radio and X-ray bands.
We then assume \citep{tanakatakahara2010, bucciantini2011, martin2012, torres2013, zhu2015}:
 \begin{equation} 
\begin{split}
Q(E,t) = Q_0(t) \begin{cases}
 (E/E_b)^{-\alpha_1} & \text{for } E \le E_b \\ 
 (E/E_b)^{-\alpha_2} & \text{for } E > E_b
 \end{cases},
\end{split}
\label{eq:injection}
\end{equation}   
where $Q_0(t)$ is a normalization factor determined from the fraction of the spin-down luminosity $L(t)$ of the pulsar that goes in particles energy and $E_b$ is the break energy where the slope of the particle spectrum changes.
If we write the spin-down luminosity of the pulsar in the form
 \begin{equation} 
\label{eq:spindown}
    L(t) = L_0\biggl(1+\frac{t}{\tau_0}\bigg)^{-\frac{(n+1)}{(n-1)}},
\end{equation}   
where $L_0$ is the initial spin-down luminosity, $\tau_0$ the initial spin-down timescale and $n$ the braking index \citep{gaensler&slane2006}, we can find the normalization factor $Q_0(t)$ from:
 \begin{equation} 
    (1-\eta_B)L(t) = \int_{E_{min}}^{E_{max}} EQ(E,t)dE.
\end{equation}   
Here the constant $\eta_B$, the magnetic fraction of the nebula \citep{martin2016}, is the fraction of the spin-down luminosity that goes into the electromagnetic field\footnote{This is not to be confused with the so called magnetization parameter $\sigma(t)=\eta_B/(1-\eta_B)$.}. While $1-\eta_B$ is the fraction of the spin-down luminosity that goes in the kinetic energy of the electrons.  

The escape timescale $\tau_{esc}(E,t)$ is computed from the assumption that particles can escape from the nebula because of diffusion. This diffusion inside a PWN arises from the interaction of particles with irregularities in the magnetic field \citep{vorster2013}. Assuming that the diffusion of plasma across the magnetic field in the PWN follows Bohm law, $\tau_{esc}(E,t)$ is given by:
 \begin{equation} 
    \tau_{esc} = \frac{e B(t)R^2_{pwn}(t)}{2Ec^2},
    \label{eq:time_escape}
\end{equation}   
where $R_{pwn}$ is the radius of the PWN.

The second term in equation \ref{eq:transport} includes the energy variation because of synchrotron radiation, IC scattering, Self-Synchrotron Compton (SSC) and adiabatic losses \citep{ginzburg&syro1964}.

The minimum energy $E_{min}$ of the injected electrons is a free parameter in this model and we choose to select a value equal to the electrons rest mass energy ($0.51$ MeV).
On the other hand the maximum electron energy $E_{max}$ has to be determined because it is strictly related to the accelerations processes at the termination shock.
There are different ways to calculate $E_{max}$.
For high magnetic field strengths (for very young PWNe) one can estimate it by balancing synchrotron losses acceleration gains \citep{dejager1996}. For lower magnetic field strengths, one needs to consider that the highest energy particle must have a gyro-radius comparable to the shock radius to participate to the acceleration process \citep{dejager&djannati2009}.
Another possibility for estimating $E_{max}$ is to consider the electric potential of the neutron star magnetosphere \citep{bandiera08, bucciantini2011, granot2017} and determine the maximum energy that electrons can gain while moving through the polar cap potential.
We computed $E_{max}$ considering all three different approaches and adopted the second one because the other two yield unreasonably high values.
The second condition is equivalent to impose that the Larmor radius $R_L$ must be a fraction $\epsilon<1$ ($\epsilon$ containment factor) of the termination shock radius $R_S$. The Larmor radius can be written as
 \begin{equation} 
    R_L = \frac{E_{max}}{eB_S},
\end{equation}   
and so the maximum energy becomes:
 \begin{equation} 
    E_{max} = \epsilon eB_SR_S.
\end{equation}   
Finally we need an expression for the magnetic field at the termination shock $B_S$. From \cite{kennel&coroniti1984} the post-shock field is expressed as:
 \begin{equation} 
    B_S = \kappa \sqrt{\eta_B \frac{L(t)}{c}} \frac{1}{R_S},
\end{equation}   
where $\kappa$ is the magnetic field compression ratio taken equal to 3 (strong shock condition).
The final expression for the maximum electron energy is then:
 \begin{equation} 
\label{eq:ele_max}
    E_{max} = 3e\epsilon\sqrt{\eta_B \frac{L(t)}{c}}.
\end{equation}   

To compute the evolution of the magnetic field we consider the adiabatic losses due to expansion work done by the nebula on the surroundings and the energy input from the pulsar wind \citep{pacini&salvato1973, torres2013, gelfand2009}:
 \begin{equation} 
    \frac{dW_B(t)}{dt} = \eta_B L(t) - \frac{W_B(t)}{R_{pwn}(t)}\frac{dR_{pwn}}{dt},
\end{equation}   
where $W_B=(4\pi/3)R_{pwn}^3(t) B^2(t)/(8\pi)$ is the total magnetic energy. The integration over time of this equation leads to
 \begin{equation} 
\label{eq:mag_evo}
    B(t) = \frac{1}{R_{pwn}^2(t)}\sqrt{6\eta_B \int_0^t L(t')R_{pwn}(t')dt'}.
\end{equation}   

The last ingredient of the model is the dynamical evolution (radius and the expansion velocity) of the PWN. We compute it with an iterative approach that is explained in appendix \ref{appendix:A}\footnote{The caveats of this iterative approach are described at the end of appendix \ref{appendix:A}.}.

The diffusion-loss equation (equation \ref{eq:transport}) is solved using a freely available code, called GAMERA\footnote{\url{libgamera.github.io/GAMERA/docs/main_page.html}} \citep{gamera2015}. 
Once the evolution of the particle spectrum is computed, it is possible to derive directly the photon spectrum with GAMERA.
The synchrotron spectrum is computed considering an isotropic pitch angle distribution of the electrons as in \cite{ghisellini1988}. The IC emission is computed using the full Klein-Nishina cross-section \citep{blumental1970} on a background radiation field (generally composed by the CMB photons and two Infra-Red components). Synchrotron Self-Compton (SSC) emission is also included  \citep{atoyan&aharonian1996}.

\subsection{Model test and comparison}
\label{sec:model_test}

The model has several parameters that constrain various physical properties of the system.
Since some of them are significantly degenerate, as the distance and the age of the system, we decide to fix them by choosing reliable value as reported in the literature (age, distance, energy of the SN explosion, density of the interstellar medium and photon background, see Table \ref{tab:model_fit}).
In addition to these parameters, several parameters of the pulsar (spin-down luminosity, period derivative, characteristic age) are also known and are reported in Table \ref{tab:model_fit}.
The remaining parameters are those related to the spectrum of the injected electrons population (the break energy and the two indices of the broken power law), the magnetic fraction of the nebula and the containment factor.

When fitting the data we leave the injection parameters free to vary. The only exceptions are $\alpha_1$ and $E_{b}$ that can be constrained from the radio and X-ray data. As already stated, changing some of the fixed parameters could, in principle, lead to very different values for the fitted parameters. For example, changing the distance of the system would lead to different values for the ejected mass of the SN, the age of the system and the densities of the background photon fields for preserving the radius and TeV flux. This would in turn lead to estimating completely different parameters for the nebula. The distance of the source must be estimated accurately to break this degeneracy.  However, once the distance is fixed at a certain value, the fitted parameters are fairly well determined. In the following we will not consider this degeneracy and we will fix the distance of the source to 13.3 kpc \citep[as reported by][]{abdalla_hess2017}, since determining it is not the main focus of this paper. The fitting procedure and error estimation of the fitted parameters are reported in Appendix \ref{appendix:B}. 

We tested our implementation against the results presented in \cite{zhu2018} and \cite{torres2014}, selecting the same set of data for consistency. The radio data are taken from \cite{2008A&A...487.1033D}, the X-ray data from \cite{2003A&A...401..197P} and the current VHE data from \cite{2005A&A...432L..25A}.  For the X-ray data, in performing the fit we considered only two points, one at the lower and the other at the higher bound of the energy interval (with the corresponding errors). They were computed from the best-fitting power law reported by \cite{2003A&A...401..197P}. The rational behind this choice was to avoid giving too much weight to the X-ray data in comparison with the radio data (with only three points) , to sample with a similar number of points the synchrotron and the IC peaks (5 and 7 points, respectively), and to comparatively increase the weight of the TeV data in the following section. This is crucial to understand to what extent the better quality of the CTA data will help in estimating the parameters of PWNe.

The values of the fitted parameters and their comparison with those found in \cite{zhu2018} and \cite{torres2014} are reported in Table \ref{tab:model_fit}.
Results are consistent. However, the (fixed) value of the ejected mass $M_{ej}$ is slightly different.
This difference is likely caused by differences in the approach adopted to solve Equation \ref{eq:transport}. However, the discrepancy does not appear to be particularly relevant considering the actual uncertainty on the knowledge of this parameter.

\begin{figure}
	\centering
	\includegraphics[width=1.\columnwidth]{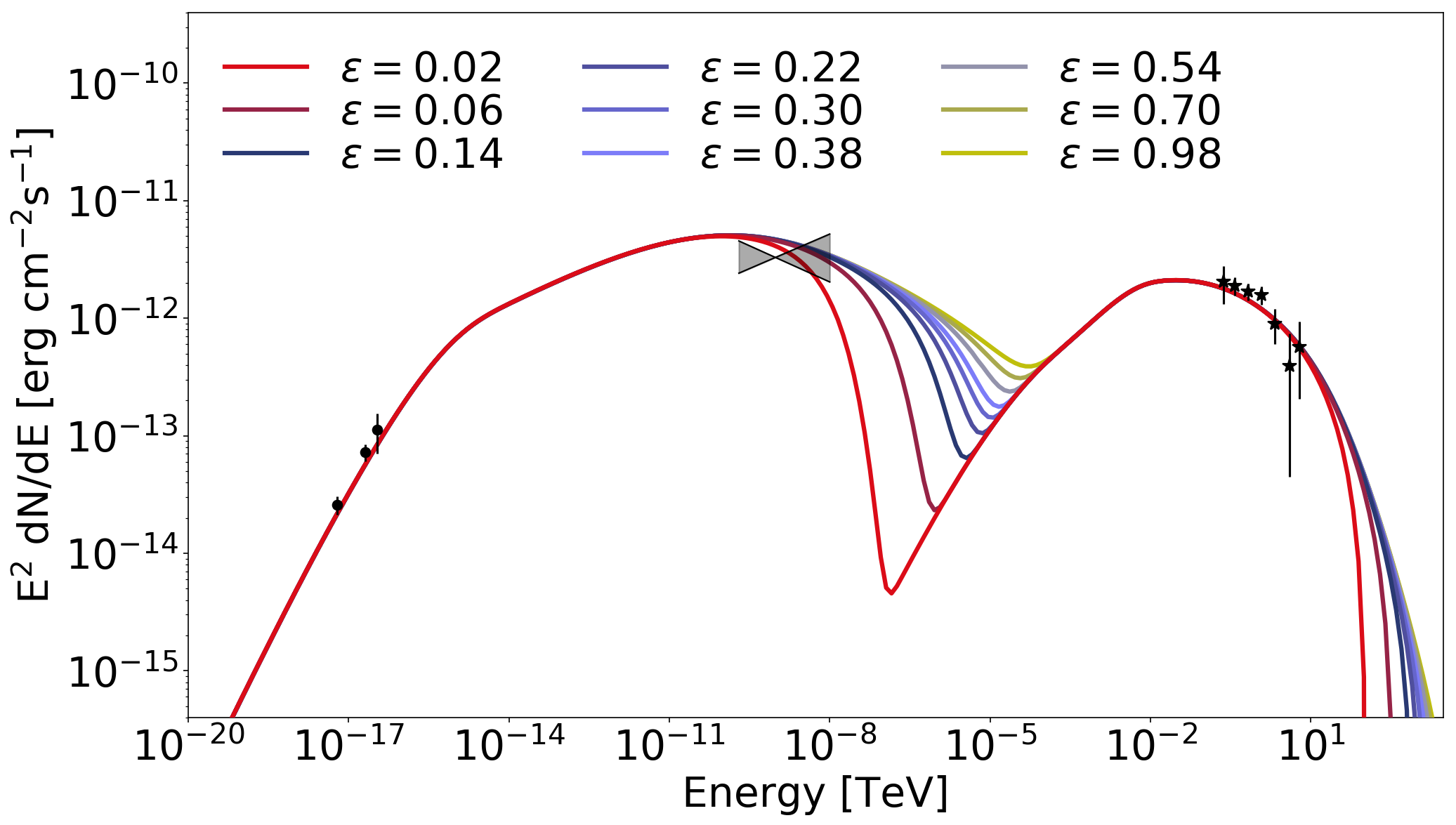}
	\caption{SED models for SNR G0.9+0.1 computed for $\alpha=2.51$ and $\eta_B = 0.031$ fixed while $\epsilon$ is varying from 0.02 to 0.98 with step 0.04. We can clearly see that is possible to rule out only really small values of the containment factor ($\epsilon\lesssim0.1$). }
	\label{epsilon}
\end{figure}

We emphasize that the parameter $\epsilon$ is loosely constrained because the data do not cover the part of the spectrum where the effects of this parameter are more evident (i.e. in the high energy tails of the synchrontron and IC peaks). Is possible to see this effect in Figure \ref{epsilon} where we vary only $\epsilon$ between $0.02$ and $0.98$ with a constant step of $0.04$. This parameter is only constrained to be $>0.1$. We then took $\epsilon=0.25$ as reference value for all the models in the subsequent analysis.

\begin{figure}
	\centering
	\includegraphics[width=1.\columnwidth]{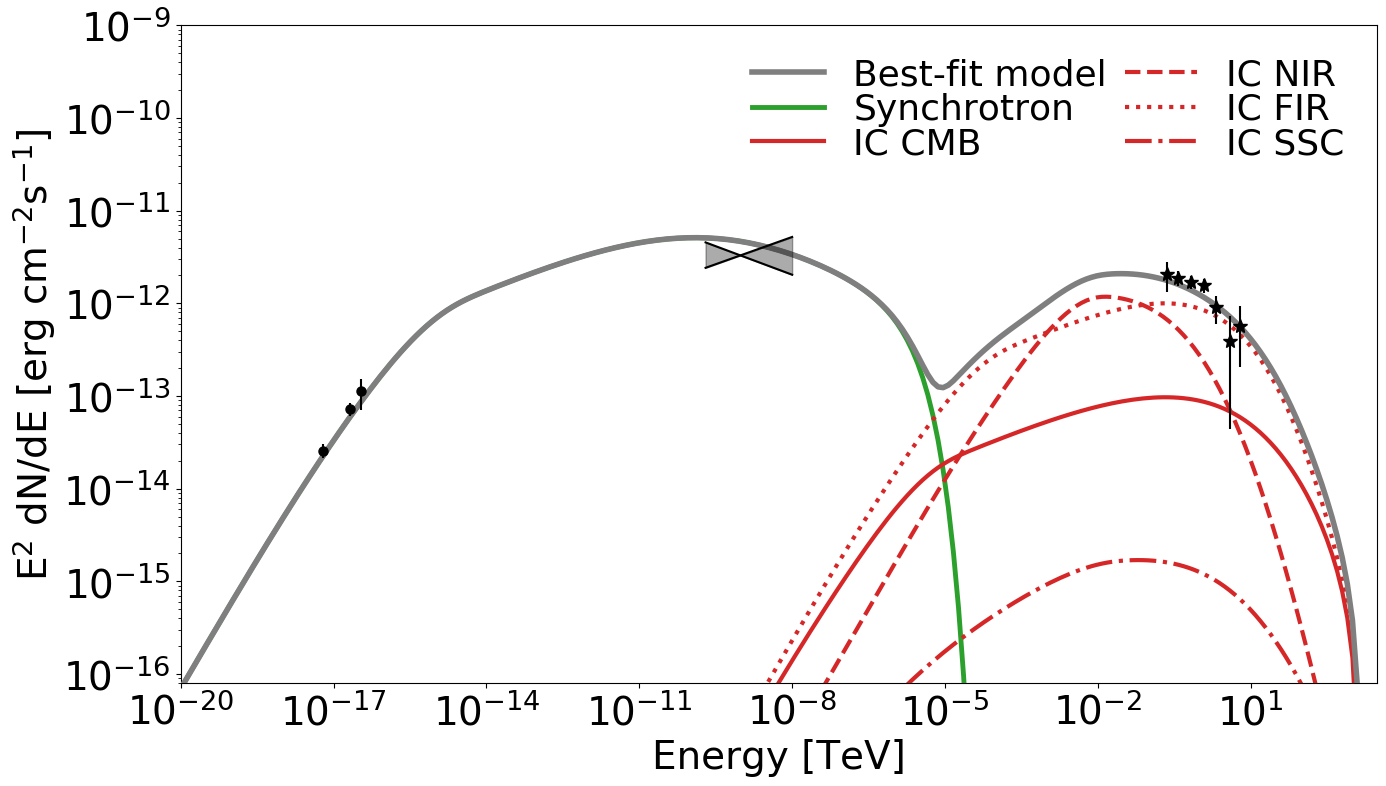}
	\includegraphics[width=1.\columnwidth]{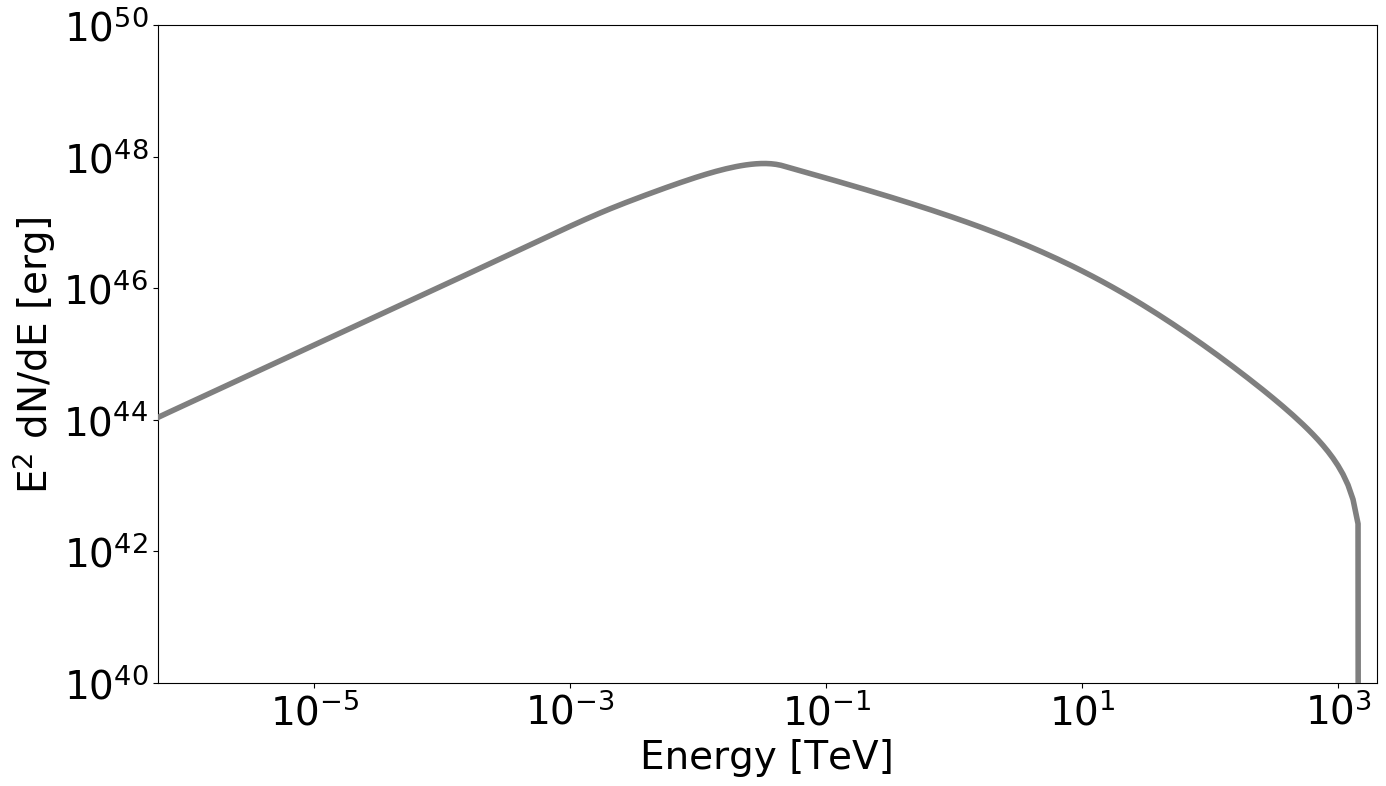}
	\caption{Photon (upper panel) and electron (lower panel) SED model for SNR G0.9+0.1 computed with the best fit values reported in Table \ref{tab:model_fit}. The dataset is the same as in the work of \protect\cite{zhu2018} (more detail in the text).}
	\label{hessfit}
\end{figure}

Figure \ref{hessfit} shows the final best fit electron and photon spectra. The reduced chi square of the fit is $\chi^2_\nu = 1.1$\footnote{The reported value of the reduced chi square is not to be intended as an absolute measurement of the goodness of the fit on the original complete data set (we did not consider all the X-ray spectral points), but only as a reference value useful for comparison with the fits of the simulated data reported below.}.

\begin{table*}
    \centering
    \caption{Fixed and fitted parameters of the model in comparison with those of \protect\cite{zhu2018} and \protect\cite[Model 2]{torres2014}. Dots mean that the same value is adopted. All the parameters are computed for the estimated age $t_{a}$.}
    \label{tab:model_fit}
    \def\arraystretch{1.2}
    \begin{tabular}{lcccc}\hline\hline
 & This work & \cite{zhu2018} & \cite{torres2014} & Notes\\\hline
\multicolumn{5}{l}{\textbf{Pulsar and SN parameters (fixed)}}\\\hline
$P$ [ms] & $52.2$ & $...$ & $...$ & from \cite{camilo2009}\\
$\dot{P}$ [s s$^{-1}$]& $1.56\times10^{-13}$ & $...$ & $...$ &  from \cite{camilo2009} \\
$\tau_c$ [yr] & $5305$ & $...$ & $...$ &  $P/(n-1)\dot{P}$ \\
$n$ & $3$ & $...$ & $...$ & fixed at the standard braking index value\\
$L(t_{a})$ [erg/s] & $4.32\times10^{37}$ & $...$ & $...$ & from \cite{camilo2009} \\
$t_{a}$ [yr] & $3000$ & $...$ & $...$ & estimated age\textsuperscript{a} \\
$\tau_0$ [yr] & $2305$ & $...$ & $...$ & $[2\tau_c/(n-1)]-t_{a}$ \\
$L_0$ [erg/s] & $2.29\times10^{38}$ & $...$ & $...$ & from equation \ref{eq:spindown}\\
$M_{ej}$ [M$_{\odot}$] & $9$ & $14$ & $17$ & estimated\textsuperscript{a}\\
$E_{sn}$ [erg] & $10^{51}$ & $...$ & $...$ & estimated\textsuperscript{a}\\
$d$ [kpc] & $13.3$ & $...$ & $13.$ & from \cite{abdalla_hess2017}\\\hline
\multicolumn{5}{l}{\textbf{Environment parameters (fixed)}}\\\hline
$n_h$ [cm$^{3}$] & $0.01$ & $...$ & $1.$ & from \cite{zhu2018} \\
$T_{CMB}$ [K] & $2.7$ & $...$ & $...$ & from \cite{LongairGalaxyForm} \\
$U_{CMB}$ [eV/cm$^3$] & $0.25$ & $...$ & $...$ & from \cite{LongairGalaxyForm}\\
$T_{FIR}$ [K] & $30$ & $...$ & $...$ & from \cite{torres2014} \\
$U_{FIR}$ [eV/cm$^3$] & $3.8$ & $...$ & $...$ & from \cite{torres2014}\\
$T_{NIR}$ [K] & $3000$ & $...$ & $...$ & from \cite{torres2014} \\
$U_{NIR}$ [eV/cm$^3$] & $25$ & $...$ & $...$ & from \cite{torres2014} \\\hline
\multicolumn{5}{l}{\textbf{Injection parameters}}\\\hline
$E_{b}$ [TeV]& $0.045$ & $...$ & $0.026$ & from \cite{zhu2018}\\
$\alpha_1$ & $1.1$ & $...$ & $1.2$ & from \cite{zhu2018}\\
$\alpha_2$ & $2.523\pm0.022$ & $2.52\pm0.02$ & $2.5$ & fitted\\
$\eta_B$ & $0.0313\pm0.0055$ & $0.029\pm0.004$ & $0.02$  & fitted\\
$\epsilon$ & $>0.10$ & $0.25\pm0.08$ & $0.2$ & fitted\\\hline
\multicolumn{5}{l}{\textbf{PWN parameters\textsuperscript{b}}}\\\hline
$R_{pwn} (t_{a})$ [pc] &$3.46\pm0.01$ & $3.51$ & $3.8$ & from iterative procedure in Appendix \ref{appendix:A}\\
$B (t_{a})$ [$\micro$G] & $21.89\substack{+1.93 \\ -2.08}$ & $20.29^{+1.86}_{-1.93}$ & $15$ & from equation \ref{eq:mag_evo}\\
$E_{max}(t_{a})$\textsuperscript{c} [TeV] & $>600$ & $1452^{+600}_{-535}$ & $971$ & from equation \ref{eq:ele_max}\\\hline
\multicolumn{5}{l}{\scriptsize{\text{\textsuperscript{a} $t_{a}$, $M_{ej}$ and $E_{SN}$ taken in order to obtain a nebula of $\sim2'$ located at 13.3 kpc. \textsuperscript{b} Computed from the PWN dynamics (see Appendix \ref{appendix:A})}}}\\
\multicolumn{5}{l}{ \scriptsize{\text{\textsuperscript{c} Maximum energy of the electrons in injection at the termination shock of the nebula.}}}\\
    \end{tabular}
\end{table*}

\subsection{Fit of simulated CTA data}

We applied this model of the PWN evolution to the various simulated spectra of SNR G0.9+0.1 reported in Section \ref{sec:simSNR}, assuming that the most of the simulated VHE emission comes from the central PWN.
The spectral range is limited at 200 GeV to be consistent with the lower limit of the HESS data and we rebin the spectrum with 10 bins. However, depending on the spectral shape, the spectrum can have less than 10 bins since at high energies there may be no photons.

\begin{figure*}
	\centering
	\includegraphics[width=1.\columnwidth]{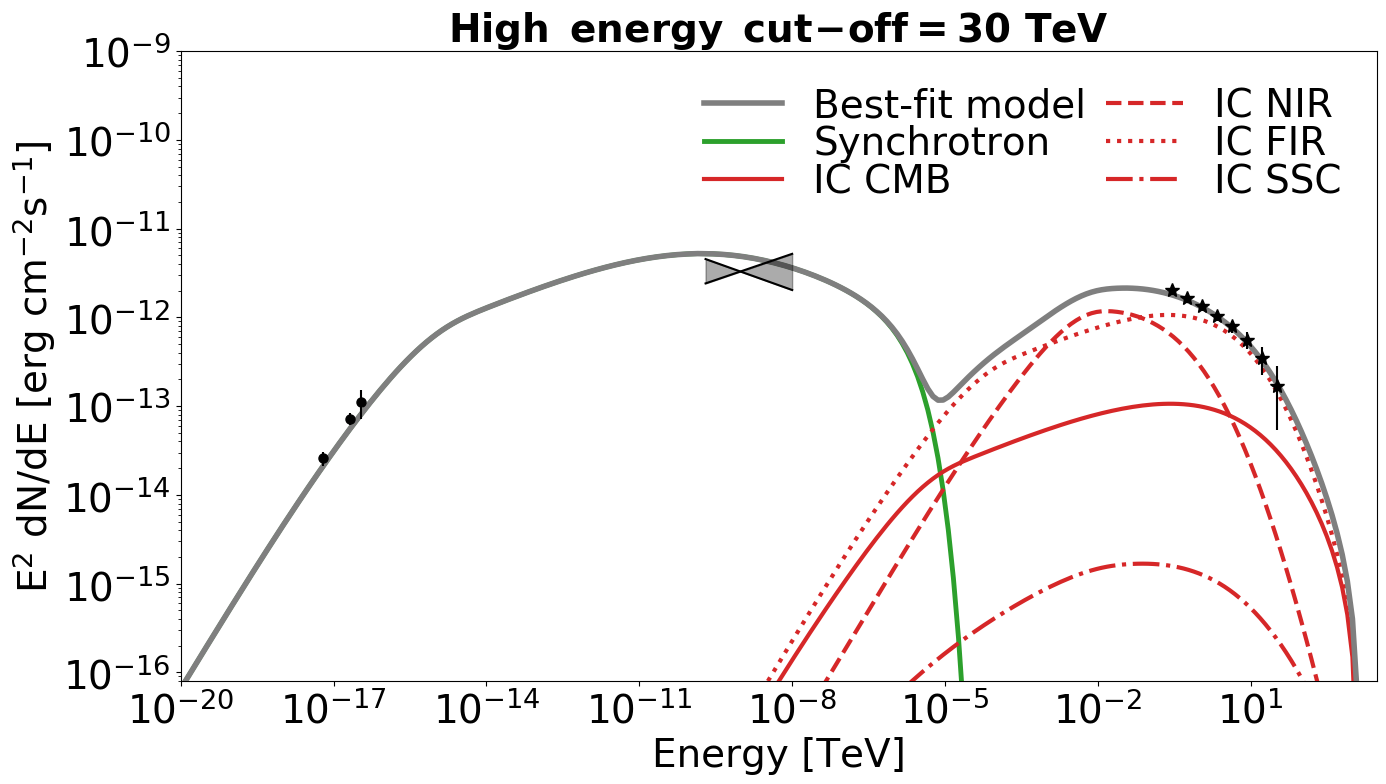}
	\includegraphics[width=1.\columnwidth]{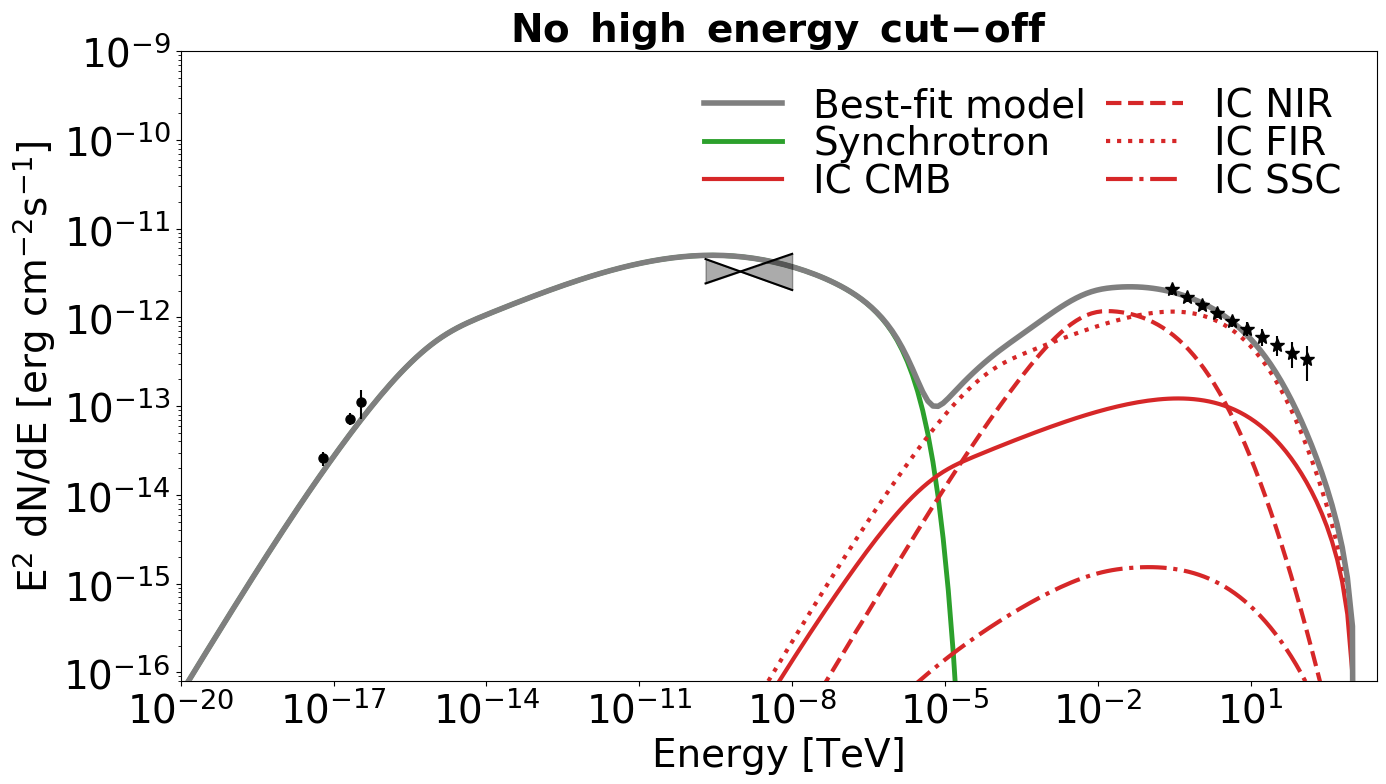}
	\caption{Photon SED computed with the best-fitting parameters for two out of the six different CTA simulated spectra of SNR G0.9+0.1 reported in Table \ref{tab:cta_model_fit}.}
	\label{ctaphysfit}
\end{figure*}

The results of the model fit are reported in Table \ref{tab:cta_model_fit}, while two representative MWL spectra are shown in Figure \ref{ctaphysfit}. The errors on the $\gamma$-ray data-points includes also the systematic errors computed in the previous section.

From Figure \ref{ctaphysfit} we see that in the lower energy part of the spectrum (the synchrotron emission peak) the model is always consistent with the data, while this is not the case at high energies.

The most interesting results is that the value of the magnetization parameters $\eta_B$ is fairly well determined and tend to decrease with increasing cut-off energy,  because, for energy conservation, increasing the maximum energy of the electrons requires that more power goes in particles ($1-\eta_B$) and less in the magnetic field. In general, the MWL spectrum, can constrain it.

For a cut-off at $20-30$ TeV we found a good agreement of the fitted parameters with the values obtained from the HESS data. For a cut-off at a different energy the inferred parameters have significantly different values, which means that with the data currently available it is not possible to accurately constrain them. With the CTA data, which have a higher energy threshold, the estimates will be more accurate.  The increased sensitivity of CTA will then allow us to observe this and other PWNe at higher energies and make accurate studies on how particles are accelerated at the termination shock.

Finally we want to emphasize that the model spectra are not consistent with a pure power law simulated spectrum for every value of the parameters (reduced $\chi^2_\nu \simeq 2.3$). 
With this model we are not able to reproduce a power law with no measured cut-off. Even changing the age and distance of the source, it is not possible to find a model that has a power law tail up to 180 TeV. The only possibility would probably be including an hadronic component, but this is beyond the purpose of this work.

\begin{table*}
    \centering
    \caption{Results of the fitting procedure with the PWN model adopted in this work for various CTA simulated spectra of SNR G0.9+0.1.}
    \label{tab:cta_model_fit}
    \def\arraystretch{1.2}
    \begin{tabular}{lcccccc}\hline\hline
 & PLEC 20 TeV & PLEC 30 TeV & PLEC 50 TeV & PLEC 80 TeV & PLEC 100 TeV & PWL \\\hline
\multicolumn{5}{l}{\textbf{Fitting result}}\\\hline
$\chi^2_\nu$ & $0.8$ & $0.8$ & $0.9$ & $1.1$ & $1.2$ & $2.3$ \\
$\alpha_2$ & $2.520\pm0.017$ & $2.509\pm0.016$ & $2.507\pm0.016$ & $2.503\pm0.016$ & $2.500\pm0.016$ & $2.496\pm0.016$\\
$\eta_B$ & $0.0310\pm0.0048$ & $0.0286\pm0.0045$ & $0.0272\pm0.0044$ & $0.0257\pm0.0042$ & $0.0251\pm0.0041$ & $0.0232\pm0.0039$\\
$R_{pwn}$ (pc) &$3.46\pm0.01$ & $3.45\pm0.01$ & $3.45\pm0.01$ & $3.45\pm0.01$ & $3.45\pm0.01$ & $3.45\pm0.01$\\
$B$ ($\micro$G) & $21.81\substack{+1.53 \\ -1.67}$ & $20.97\substack{+1.55 \\ -1.59}$ & $20.44\substack{+1.49 \\ -1.58}$ & $19.87\substack{+1.41 \\ -1.55}$ & $19.65\substack{+1.40 \\ -1.52}$ & $18.89\substack{+1.39 \\ -1.50}$ \\\hline
    \end{tabular}
\end{table*}

\subsection{Impact of the ISRF}

We now try to estimate the impact on our results caused by the uncertainties on the Inter-Stellar Radiation Field (ISRF) at the (unknown) position of SNR G0.9+0.1. In principle a different ISRF can affect our measurement of the parameter of the nebula since the shape of the IC component is dependent on the background radiation.
In the previous analysis we fixed the parameters of the ISRF. It would have been computationally too expensive to let them free.

The density and temperature of the Near-Infrared (NIR) and Far-Infrared (FIR) photon field can vary significantly with the position in the galaxy.
Moreover the spectral shape of this emission can be very different from the simple sum of diluted black-bodies (as assumed in the previous sections).

\begin{figure*}
	\centering
	\includegraphics[width=1.\columnwidth]{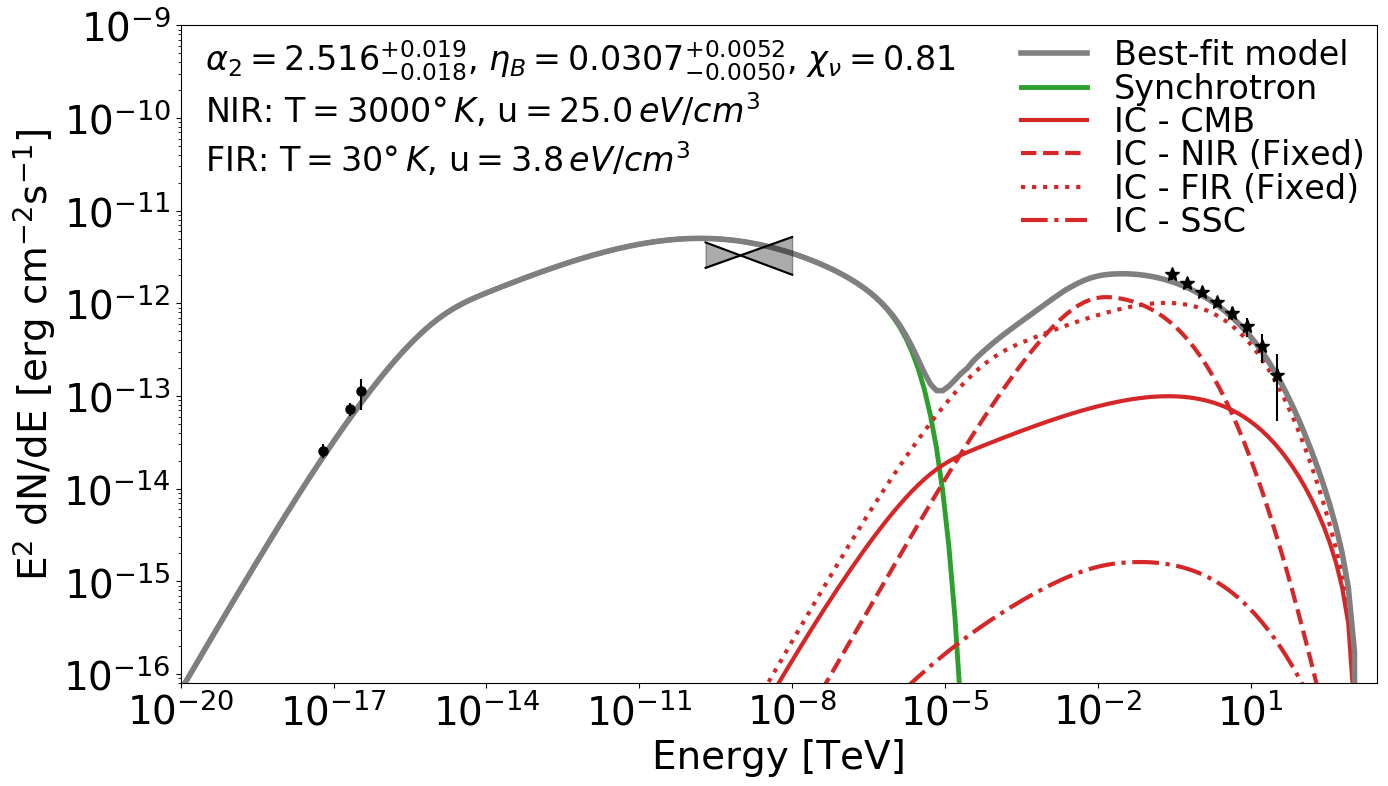}
	\includegraphics[width=1.\columnwidth]{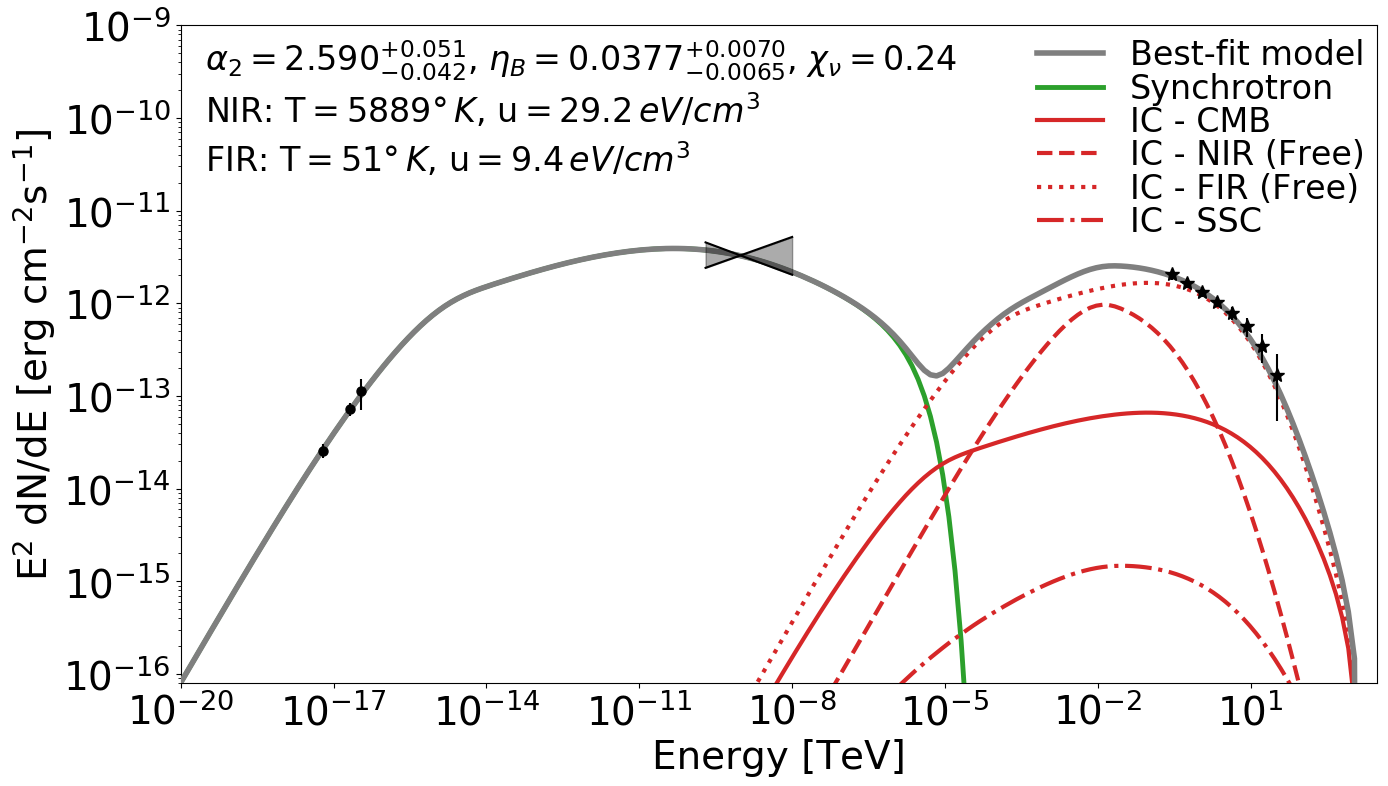}
	\caption{Photon SED computed with the fixed ISRF background (left panel) and with the free ISRF background (right panel). The best fit value are computed with an MCMC procedure.}
	\label{fig:fixedVSfree}
\end{figure*}

In order to estimate the effects of different ISRFs, we perform two different approaches. 
In the first, we check how much the fit differs comparing the case with fixed and free ISRF parameters.
To do it we cannot use the full model since the computational time would be too large.
We then treated the dynamical evolution in a simplified way, assuming a PWN freely expanding in the SNR using just equation \ref{eq:radius_simply}.
We then considered the CTA simulated data with a cut-off at 30 TeV and fitted them leaving $\alpha_2$ and $\eta_B$ free. 
We used a Monte Carlo Markov Chain (MCMC) code \citep[\textit{emcee},][]{emcee} and made 2500 realizations of the spectrum.
We obtain results similar to those previously found ($\alpha_2=2.516\substack{+0.019 \\ -0.018}$, $\eta_B=0.0307\substack{+0.0052 \\ -0.0050}$).
After this, we repeated the fit but adding as free parameters the energy density and temperature for the IR radiation fields ($T_{FIR}\,, U_{FIR}\,,T_{NIR}\,, U_{NIR}$). We found in this case a different ISFR, with an higher energy density of the Far IR component (see Figure \ref{fig:fixedVSfree}). However the relevant parameters of the PWN did not change significantly, although their errors increased ($\alpha_2=2.593\substack{+0.049 \\ -0.041}$, $\eta_B=0.0378\substack{+0.0075 \\ -0.0068}$). 

\begin{figure*}
	\centering
	\includegraphics[width=1.\columnwidth]{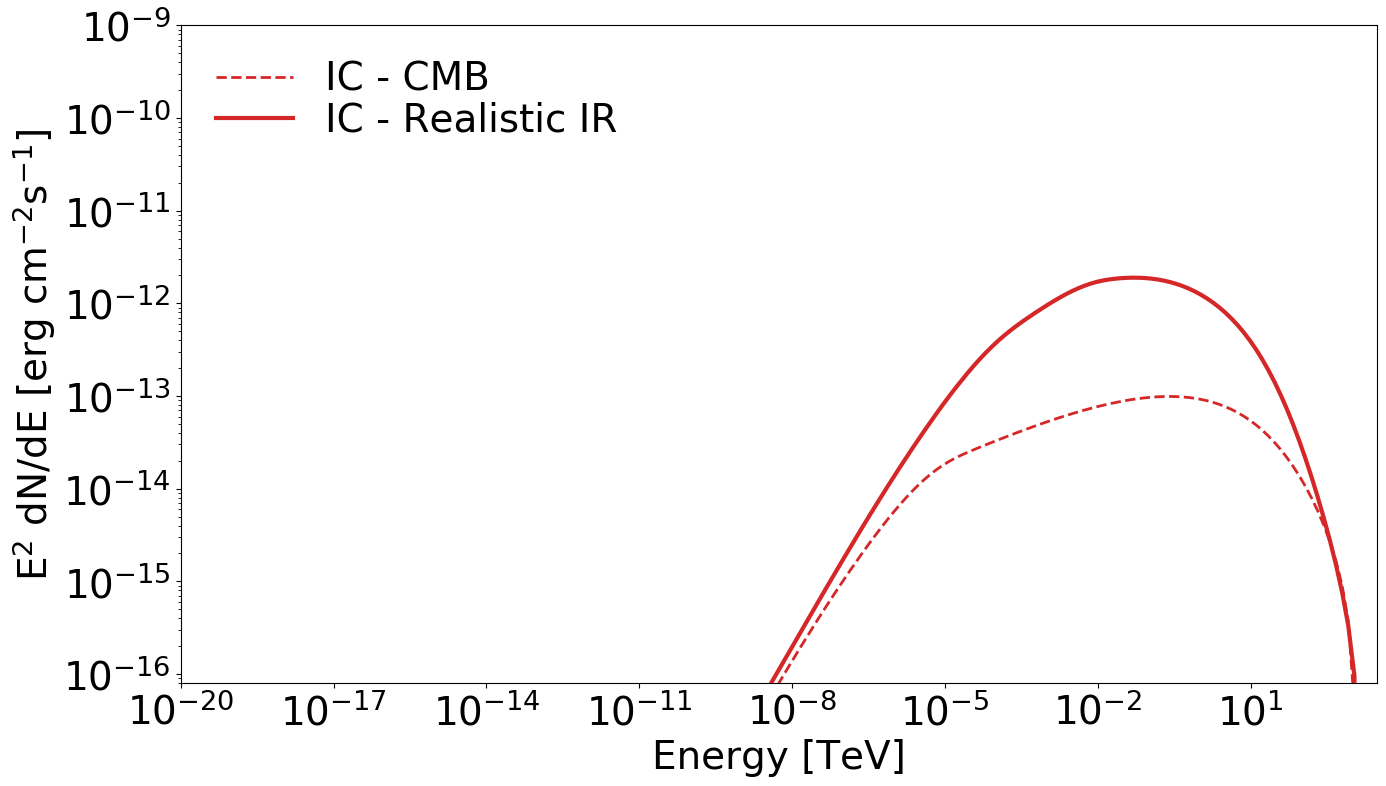}
	\includegraphics[width=1.\columnwidth]{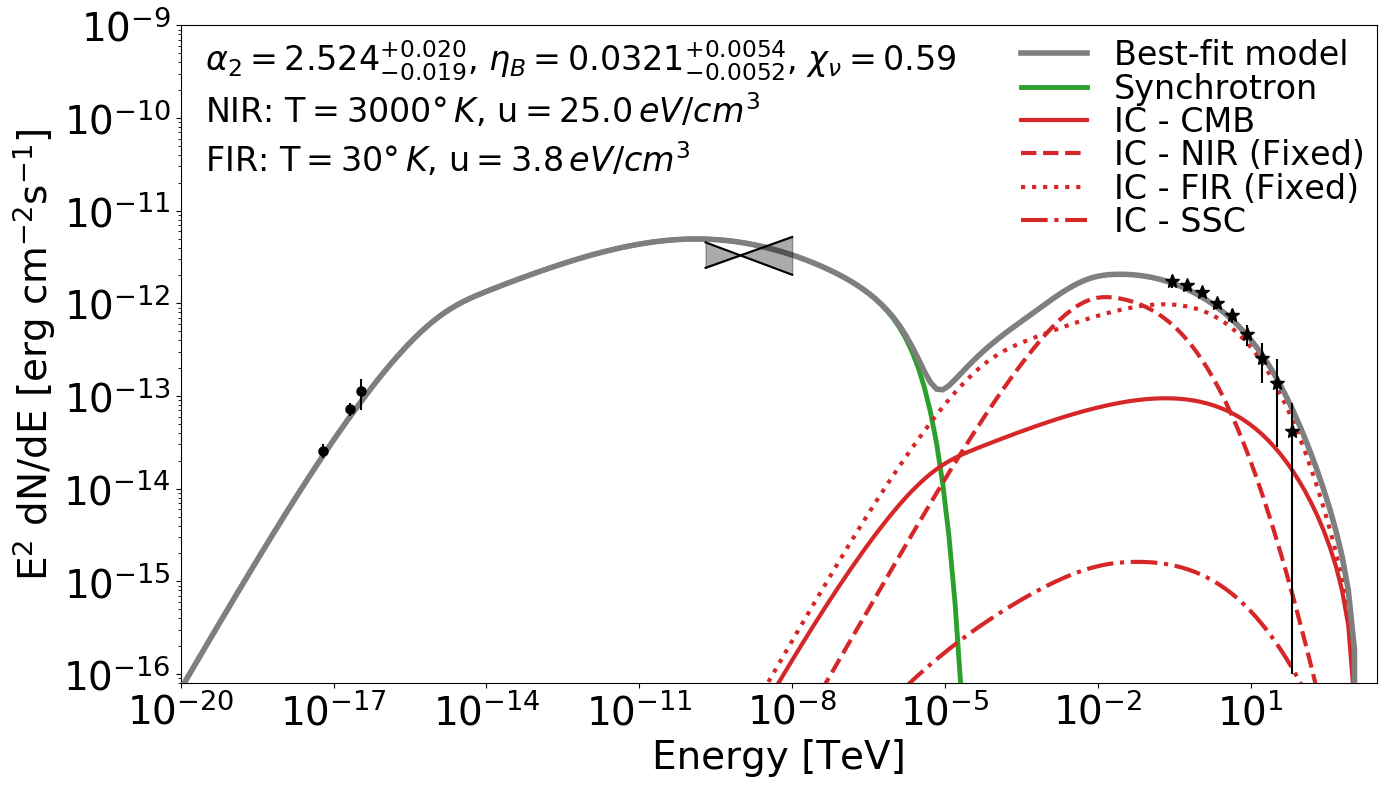}
	\caption{In the left panel we show the realistic \protect\citep{popescu2017} IR background reprocessed via IC used as input for the simulation. In the right panel the photon SED fitted with just two diluted blackbody.}
	\label{fig:realisticBKG}
\end{figure*}

In the second approach we considered  a more realistic radiation field, like the axisymmetric solution for the ISFR of the Milky Way provided by \cite{popescu2017}, and use it to produce a model with fixed nebula parameters ($\alpha_2=2.515$, $\eta_B=0.0315$). We selected the model reported in the first panel in Figure 9 of \cite{popescu2017} and rescaled it by a factor $\sim3$ to obtain a similar $\gamma$-ray flux as the one of SNR G0.9+0.1. 
We then used this model to simulate an observation made with CTA, extracted the new spectrum and used it in the MCMC fitting procedure as before. We fit the usual two parameters $\alpha_2$ and $\eta_B$ fixing again the values for the ISRF as in the previous analysis and using two diluted blackbodies to model it.
We obtained values that are in very good agreement with the ones used for the preparation of this model ($\alpha_2=2.524\substack{+0.020 \\ -0.019}$, $\eta_B=0.0321\substack{+0.0054 \\ -0.0052}$). The results are shown in Figure \ref{fig:realisticBKG}. We also tried to fit this model leaving all the parameters for the IR radiation field free to vary and found similar values.
While the energy density of the ISRF is crucial to reproduce the IC component in the VHE spectrum, its actual spectral distribution is not, because Comptonized IR photons tends to loose rapidly memory of their initial energy.

\section{Conclusions}
\label{sec:conclusion}

In this work we have studied a young PWN inside SNR G0.9+0.1 that is projected near the Galactic Center. Despite the high background rate, the crowded field and the faintness of the source we have shown that the CTA-South array enables us to study this region, and in particular the PWN, in great detail.

In our analysis of SNR G0.9+0.1 we choose 200 hours as observing time for the simulations in order to obtain very accurate data. This observing time is early achievable because of the projected position of this source, close to the Galactic Center. As reported in the book "Science with the Cherenkov Telescope Array" \citep{scienceCTA}, the Galactic Center is one of the Key Science Project (KSP) for the CTA collaboration.
This core programme will run for the first 3 years of observations with CTA and will produce 525 hours of data from the region of SNR G0.9+0.1. The 200 hours of time needed for our study will be reached after  $\sim1$ year after the beginning of the observations with CTA-South.

Our spatial analysis of SNR G0.9+0.1 shows that if the VHE emission region is larger than $\sim 0.65$ arcmin CTA will be able to resolve it, leading to a measurement of the size of the nebula in the VHE band. Furthermore our spectral analysis shows that it would be possible to distinguish different spectral models and calculate the cut-off energy, if present.
We could also detect the source at energies higher then 100 TeV if the spectrum is a pure power law.

We performed also a detailed analysis of the systematic errors and found that the systematics related to the instrumental uncertainties dominate, especially at low energies. Despite these errors maybe somewhat overestimated, they provide at least an order of magnitude estimate of the uncertainties that is crucial for our subsequent analysis.

We have then implemented a one-zone time dependent leptonic model that computes the evolution of a young PWN inside a SNR in order to obtain some physical information and to understand what impact on our knowledge of this PWN CTA may have. We first compared our result with those obtained by \cite{zhu2018} and \cite{torres2014} using the same data-set. We find good agreement, although it is difficult to constraint the confinement factor $\epsilon$ (hence the maximum energy of the injected electrons in the nebula). Measurements of the flux of SNR G0.9+0.1 at MeV energies would be needed to obtain a precise value for this parameter. However, in the absence of MeV data, an increase of the VHE observing time would help to put constraints on the maximum electron energy because the tail of the IC peak is also sensitive to it at high energies. From the best fit model of the currently available data (Table \ref{tab:model_fit}) we expect an high energy cut-off between 20 and 30 TeV. This is a measurement that CTA could easily do, as shown in Figure \ref{spectrapwl},  thus allowing us to reduce the uncertainties on the estimated parameters of the PWN (see Table \ref{tab:cta_model_fit}).

It is worth nothing that the possibility to put a constraint on the size of the VHE emission region with CTA would be crucial to check the goodness of the model, because we could compare it with the model computed radius and to the size observed at other wavelength.

We have shown that MWL data, including CTA data (that will be capable to constrain the cut-off, if present), will lead to a more precise measurement of the magnetization parameter $\eta_B$ of the nebula, that, for simplicity, we considered to be constant in space and time during the evolution of the nebula.
We note also that, with this model, it is not possible to reproduce a pure power law spectrum. If detected with CTA, this would require a drastic change in the model, such as the introduction of an hadronic component.

We checked also the effects caused by uncertainties on the ISRF field. A fit leaving the ISRF parameters free leads only to small differences in the values of $\alpha_2$ and $\eta_B$.
Even approximating a realistic ISRF with only two diluted blackbodies, the values of $\alpha_2$ and $\eta_B$ are not significantly affected.

\section*{Acknowledgements}

We thank the referee for his useful comments. We would also like to thanks to developer of the software GAMERA, J. Hahn, for the availability at answering very quickly and clearly at our questions on the usage of the code and also for making available it to all.
This paper has gone through internal review by the CTA Consortium. We made use of ctools, a community-developed analysis package for Imaging Air Cherenkov Telescope data. ctools is based on GammaLib, a community-developed toolbox for the high-level analysis of astronomical gamma-ray data.
This research made use also of the following PYTHON packages: MATPLOTLIB \citep{matplotlib}, NUMPY \citep{numpy}, ASTROPY \citep{astropy} and EMCEE \citep{emcee}.
We acknowledge financial contribution from INAF through grant "ASTRI/CTA Data Challenge (ACDC).

\section*{Data Availability}

The data underlying this article will be shared on reasonable request to the corresponding author.

\bibliographystyle{mnras}

\bibliography{g0901}



\appendix

\section{General form of energy diffusion equation}
\label{appendix:0}
Here we describe in detail the energy-diffusion equation used in this work, starting from its general, non-simplified form \citep{ginzburg&syro1964}:
 \begin{equation} 
\label{eq:transp_general}
\begin{split}
    \frac{\partial N_i(E,\vec{r},t)}{\partial t} & =  \nabla \cdot [D_i(E,\vec{r},t)\nabla N_i(E,\vec{r},t)] - \frac{\partial}{\partial E} [b_i(E) N_i(E,\vec{r},t)] \\ 
     & + \frac{1}{2}\frac{\partial^2}{\partial E^2}[d_i(E)N_i(E,\vec{r},t)] + Q_i(E,\vec{r},t) \\ 
     & - \frac{N_i(E,\vec{r},t)}{\tau_i(E,\vec{r},t)} + \sum_k \int P_i^k(E', E)N_k(E',\vec{r},t)dE,
\end{split}
\end{equation}   
$N_i(E,\vec{r},t)$ is the number density of particles species denoted with the subscript $i$.
The first term on the right-hand side describes the spatial diffusion of particles inside the nebula and $D_i(e,\vec{r},t)$ is the diffusion coefficient. The second term describes the continuous energy variation due to acceleration processes and energy losses, including adiabatic, synchrotron and IC losses. The function $b_i(E)$ is the mean energy variation of the particle in unit time.
The third term is related to fluctuations in this continuous variation of energy of the particles, whereas the function $d_i(E)$ is equal to the mean square of the energy variation per unit time. The term $Q_i(E,\vec{r},t)$ is the particle injection rate, which in this case originates from the acceleration of the particles at the termination shock.
The fifth term accounts for the escape of particles from the system with the characteristic timescale $\tau_i(E,\vec{r},t)$. 
Finally, the last term accounts for the creation and annihilation of particles with a probability distribution $P_i^k(E', E)$ \citep{ginzburg&syro1964}.

The equation \ref{eq:transp_general} cannot be easily solved. Suitable approximations are usually made. First of all, we consider only one population of particles (electrons), we neglect pair creation or annihilation and we take only the mean value of the energy losses per unit energy, neglecting any fluctuations in the continuous energy variation. We also assume an isotropic distribution of electrons, an isotropic injection term inside the nebula and a uniform magnetic field (no diffusion effect inside the nebula). With these approximations we can neglect the first, the third and the last term in equation \ref{eq:transp_general}, that becomes equation \ref{eq:transport} from Section \ref{sec:model_detail}. The escape term in equation \ref{eq:transp_general} is retained, even if we neglect the other diffusive terms. Therefore, particles are allowed to escape from the nebula, although we do not treat in detail the diffusion process.

\section{Radius and velocity evolution of PWN}
\label{appendix:A}

In this appendix we describe an iterative method similar to that from \cite{gelfand2009}, which we use to compute the radius $R_{pwn}$ and the expansion velocity $v_{pwn}$ of the PWN in each time step. For this, it is necessary to take into account an interaction between the SNR and the PWN expanding inside it. 

First of all, we determine the properties of the ejected material between the reverse shock of the remnant and the nebula. Making a standard assumption that an inner core with initially constant density is surrounded by an outer envelope with density proportional to $r^{−9}$ \citep{blondin2001, truelove&mckee1999, gelfand2009}, the density of the ejecta can be written as:
 \begin{equation} 
\label{eq:r01}
    \rho_{ej}(r,t) = \begin{cases}
        \frac{10}{9\pi}E_{sn}v_t^{-5}t^{-3} & \text{for } r\leq v_t t \\
        \frac{10}{9\pi}E_{sn}v_t^{-5}t^{-3}(\frac{r}{v_t t})^{-9} & \text{for } r>v_t t
    \end{cases},
\end{equation}   
where $v_t=(40E_{sn}/18M_{ej})^{1/2}$ is the transition velocity between the constant density core and the outer envelope, $E_{sn}$ is an energy of the supernova explosion and $M_{ej}$ is its ejected mass. The ejecta during this stage is expanding ballistically and, therefore, its velocity is equal to $v_{ej}=r/t$. Since in this work we study young PWNe, which have not reach the reverse shock of the SNR yet, we are not aiming in further modelling of the ejecta.

We adopt a thin-shell approximation \citep{chevalier2005}, considering that the expanding PWN is surrounded by a thin shell of swept-up material. 

Initial condition for our iterative procedure, which estimates the radius and the associated expansion velocity, are determined as described below. Considering the standard approximation of an isobaric bubble inside the thin-shell, where the adiabatic losses are dominant, the equation of motion of the mass of the shell $M_s$ can be written as \citep{ostriker&gunn1971, chevalier1977}
 \begin{equation} 
\label{eq:r02}
    M_{s}\frac{d^2 R}{dt^2} = 4\pi R_{pwn}^2\left[P_{pwn}-P_{ej}-\rho_{ej}\left(\frac{dR_{pwn}}{dt}-v_{ej}\right)^2\right],
\end{equation}   
where $\rho_{ej}$, $v_{ej}$ and $P_{ej}$ are computed at $R_{pwn}$, and $P_{pwn}$ is the pressure inside the nebula. 
Since in this phase $P_{pwn} \gg P_{ej}$, we can simplify neglecting the second term in the right hand side of the equation.
From the first law of thermodynamics we can write the following expression:
 \begin{equation} 
\label{eq:r03}
    \frac{dE_{pwn}}{dt} = L(t) - 4\pi P_{pwn} R_{pwn}^2 \frac{dR_{pwn}}{dt}.
\end{equation}   
This equation is possible to solve in the approximation of $t_0\ll\tau_0$ where $L(t_0)\simeq L_0$. Putting together equations \ref{eq:r01}, \ref{eq:r02} and \ref{eq:r03}, we obtain the following initial condition for the radius and expansion velocity \citep{chevalier1977, blondin2001}:
 \begin{equation} 
\label{eq:radius_simply}
    R_{pwn}(t_0) = 1.44 \left(\frac{E_{sn}^3 L_0^2}{M_{ej}^5}\right)^{1/10} t_0^{6/5},
\end{equation}   
 \begin{equation} 
    v_{pwn}(t_0) \equiv \frac{dR_{pwn}}{dt}(t_0) = \frac{6}{5}\frac{R_{pwn}(t_0)}{t_0}.
\end{equation}   

With this initial condition we can start the iterations, computing new radius of the PWN ($R_{pwn}(t+\Delta t)$) together with the magnetic field in the nebula $B_{pwn}(t + \Delta t)$ (equation \ref{eq:mag_evo}), the spin-down luminosity $L(t + \Delta t)$ (equation \ref{eq:spindown}), the maximum energy of the electrons $E_{max}(t + \Delta t)$ (equation \ref{eq:ele_max}), and the density and the velocity of the ejecta at $R_{pwn}(t + \Delta t)$.
 \begin{equation} 
    R_{pwn}(t+\Delta t)=R_{pwn}(t) + v_{pwn}(t)\Delta t.
\end{equation}   

As a second step, we computed the pressure inside the nebula, in order to determine the force acting on the shell and, therefore, a new value of the expansion velocity of the PWN. 
The net force which affects the shell is proportional to the difference between the pressure inside $P_{pwn}$ and outside the nebula $P_{ej}$:
 \begin{equation} 
    F_{pwn} \equiv \frac{d}{dt}(M_s v_{pwn})= 4\pi R_{pwn}^2 (P_{pwn}-P_{ej}).
\end{equation}   
However, the second term of this expression can be neglected since it is expected that $P_{pwn} \gg P_{ej}$.

The total pressure inside the nebula is determined as a sum of the pressure of the magnetic field $P_{pwn,B}$ and that of the moving electrons $P_{pwn,e}$. Calculating the value of the magnetic field $B_{pwn}$ from equation \ref{eq:mag_evo}, we can determine the energy stored in the magnetic field: 
 \begin{equation} 
\label{eq:EB}
    E_{pwn,B}(t) = \left(\frac{B_{pwn}^2(t)}{8\pi}\right)\frac{4\pi}{3}R_{pwn}^3(t).
\end{equation}    
From equation \ref{eq:EB} we obtain $P_{pwn,B}$ as:
 \begin{equation} 
    P_{pwn,B}(t) = \frac{E_{pwn,B}(t)}{\frac{4\pi}{3}R^3_{pwn}(t)} =\frac{B_{pwn}^2(t)}{8\pi}.
\end{equation}   

The contribution of the second component $P_{pwn,e}$ can be computed solving equation \ref{eq:transport} and extracting the total energy from the spectrum of evolved particles:
 \begin{equation} 
    E_{pwn,e}(t) = \int_{E_{min}}^{E_{max}}EN(E,t)dE.
\end{equation}   
Then, the electron pressure is found as follows: 
 \begin{equation} 
    P_{pwn,e}(t) = (\gamma_{pwn}-1)\frac{E_{pwn,e}(t)}{\frac{4\pi}{3}R^3_{pwn}(t)} =\frac{E_{pwn,e}(t)}{4\pi R^3_{pwn}(t)},
\end{equation}   
where $\gamma_{pwn}$ is equal to 4/3.

Finally, we are able to compute new expansion velocity of the nebula.
If $v_{pwn}(t)>v_{ej}(t)$ the new mass of the shell becomes
 \begin{equation} 
    M_s(t+\Delta t) = M_s(t) + \frac{4\pi}{3}\left[R_{pwn}^3(t+\Delta t)-R_{pwn}^3(t)\right]\rho_{ej}(t+\Delta t).
\end{equation}   
Otherwise, new mass $M_s(t+\Delta t)$ is simply equal to $M_s(t)$.
The new velocity $v_{pwn}(t +\Delta t)$, which will be used for calculating the radius of PWN in the next iteration, can be found from the following expression:
 \begin{equation} 
    v_{pwn}(t+\Delta t) = \frac{M_s(t)v_{pwn}(t)+\Delta M_s 
    v_{ej}(t)+F_{pwn}(t)\Delta t}{M_s(t+\Delta t)},
\end{equation}   
where $\Delta M_s = M_s(t+\Delta t)-M_s(t)$.

To compute an evolution of leptons using this iterative procedure, we solve advective equation \ref{eq:transport} many times. In case of high energy losses these computations can become time consuming. To speed up the calculations, we put an upper limit on the magnetic field inside the nebula during the first stages of evolution of the system.
We impose that magnetic field does not exceed $2000\,\micro G$ during the first 5 yrs and it is $<200\,\micro G$ up to 500 yrs of evolution.
These constrains introduce modest impact to the calculation of the radius of the source. Resulting value of the radius is $<5\%$ higher than that computed with no upper limits on the magnetic field. It is worth to mention that this approximation has been tested only for SNR G0.9+0.1 and may not be valid for younger sources (less than $\sim1000$  years), where an higher threshold for the magnetic field will be probably needed to better reproduce the observed data.
We finally note that once the values needed to determine an evolution of the nebula are obtained, we recalculate the particle spectrum without any limit on the magnetic field. We also checked that the final photon spectrum does not differ significantly from that obtained using no upper limits on the magnetic field.

\section{Model fitting}
\label{appendix:B}

In our fitting procedure we first compute a grid of models spanning a large range of values of free parameters. We then compute the chi-square $\chi^2$ statistics for each model of the grid and the observational data, and choose the best-fit model with the minimal $\chi^2$. As mentioned in Section \ref{sec:model_test}, we leave free to vary only 3 parameters: $\alpha_2$, $\eta_B$ and $\epsilon$. Other two parameters $E_b$ and $\alpha_1$ are fixed to values as in \cite{zhu2018} in order to perform comparison with their results. Finally, we estimate uncertainties of free parameters using the following procedure:
\begin{itemize}
    \item We produce a three-dimensional (3D) probability grid from the $\chi^2$ values obtained for all the models:
     \begin{equation} 
    P_{3D}(\alpha_2, \eta_B, \epsilon) \propto \exp{\left(-\chi^2/2\right)},
    \end{equation}   
    \item and normalize it:
     \begin{equation} 
    \sum_{\alpha_2, \eta_B, \epsilon} P_{3D}(\alpha_2, \eta_B, \epsilon) = 1,
    \end{equation}   
    \item We then extract the marginalized (1D) probability distribution for each parameter summing over other two parameters:
     \begin{equation} 
        P_{1D}(\alpha_2) = \sum_{\eta_B, \epsilon} P_{3D}(\alpha_2, \eta_B, \epsilon) ,
    \end{equation}   
     \begin{equation} 
        P_{1D}(\eta_B) = \sum_{\alpha_2, \epsilon} P_{3D}(\alpha_2, \eta_B, \epsilon) ,
    \end{equation}   
     \begin{equation} 
        P_{1D}(\epsilon) = \sum_{\alpha_2, \eta_B} P_{3D}(\alpha_2, \eta_B, \epsilon) .
    \end{equation}   
    \item Finally, using these marginalized probability distributions, we estimate the confidence interval and $1\sigma$ error for each parameter, assuming that the distributions are Gaussians.
\end{itemize}
\label{lastpage}
\end{document}